\documentclass[final]{cvpr}

\usepackage{times}
\usepackage{epsfig}
\usepackage{graphicx}
\usepackage{amsmath}
\usepackage{amssymb}

\usepackage[ruled,vlined]{algorithm2e}
\usepackage{subcaption}
\usepackage{mathtools}
\usepackage{cite}
\usepackage{tabularx}
\usepackage{soul}
\usepackage{comment}
\usepackage[acronym]{glossaries}
\usepackage[percent]{overpic}
\usepackage[dvipsnames]{xcolor}
\usepackage{pifont}

\DeclarePairedDelimiter{\floor}{\lfloor}{\rfloor}
\graphicspath{ {./images} }

\newcommand*\colourcheck[1]{%
  \expandafter\newcommand\csname #1check\endcsname{\textcolor{ForestGreen}{\ding{52}}}%
}

\newcommand*\colourmark[1]{%
  \expandafter\newcommand\csname #1mark\endcsname{\textcolor{#1}{\ding{55}}}%
}
\colourmark{red}
\colourcheck{blue}
\colourcheck{green}
\colourcheck{red}

\makeatletter

\renewcommand\subsubsection{\@startsection{subsubsection}{3}{\z@}%
                                     {-1ex\@plus -1ex \@minus -.2ex}%
                                     {-1.5ex \@plus -.2ex}
                                     {\normalfont\normalsize\bfseries}}
\makeatother

\usepackage[pagebackref=true,breaklinks=true,colorlinks,bookmarks=false]{hyperref}

\newacronym{ct}{CT}{Computed Tomography}
\newacronym{fbp}{FBP}{Filtered Back-Projection}
\newacronym{dp}{DP}{Discriminator Perceptual}
\newacronym{gan}{GAN}{generative adversarial networks}
\newacronym{vgg}{VGG}{Visual Geometry Group}
\newacronym{ir}{IR}{Iterative Reconstruction}
\newacronym{tv}{TV}{total variation}
\newacronym{sin}{SIN}{Sinogram Inpainting Network}
\newacronym{prn}{PRN}{Postprocessing Refining Network}
\newacronym{pp}{P\&P}{Plug-and-Play}
\newacronym{mar}{MAR}{metal artifact removal}
\newacronym{obl}{OBL}{operator-based learning}
\newacronym{sv}{SV}{sparse-view}
\newacronym{svct}{SV-CT}{Sparse-View CT}
\newacronym{mlp}{MLP}{multi-layer perceptron}
\newacronym{psnr}{PSNR}{Peak Signal-to-Noise Ratio}
\newacronym{ssim}{SSIM}{Structural Similarity Index Measure}
\newacronym{cnn}{CNN}{convolutional neural network}
\newacronym{pn}{PN}{postprocessing networks}

\begin{document}

\title{2-Step Sparse-View CT Reconstruction with a Domain-Specific Perceptual Network}

\author{Haoyu Wei$^{1}$\thanks{The first two authors have equal contribution.}
\qquad
Florian Schiffers$^{1}$\footnotemark[1]
\qquad
Tobias Würfl$^{2}$
\qquad
Daming Shen$^{3}$\\
\qquad
Daniel Kim$^{3}$
\qquad
Aggelos Katsaggelos$^{1}$
\qquad
Oliver Cossairt$^{1}$\\
$^{1}$Northwestern University, Evanston, USA
\qquad
$^{2}$University of Erlangen-Nuremberg, Germany\\
\qquad
$^{3}$Feinberg School of Medicine, Northwestern University, Chicago, USA\\
\qquad
{\tt\small florian.schiffers@northwestern.edu}
}

\maketitle

\begin{abstract}
   Computed tomography is widely used to examine internal structures in a non-destructive manner.
   To obtain high-quality reconstructions, one typically has to acquire a densely sampled trajectory to avoid angular undersampling.
   However, many scenarios require a sparse-view measurement leading to streak-artifacts if unaccounted for.
   Current methods do not make full use of the domain-specific information, and hence fail to provide reliable reconstructions for highly undersampled data.
   
   We present a novel framework for sparse-view tomography by decoupling the reconstruction into two steps: 
   First, we overcome its ill-posedness using a super-resolution network, SIN, trained on the sparse projections.
   The intermediate result allows for a closed-form tomographic reconstruction with preserved details and highly reduced streak-artifacts.
   Second, a refinement network, PRN, trained on the reconstructions reduces any remaining artifacts.
   
   We further propose a light-weight variant of the perceptual-loss that enhances domain-specific information, boosting restoration accuracy.
   Our experiments demonstrate an improvement over current solutions by $4$ dB.
\end{abstract}

\section{Introduction}

Since invented in 1972, \gls{ct} has quickly been recognized as an essential non-destructive modality for industrial inspection~\cite{de2014industrial}, medical diagnosis~\cite{maier2018medical} or material sciences~\cite{withers2007x}.
\gls{ct} is a textbook example of computational imaging, where a physical quantity is encoded in the measured data by an image formation model and then reconstructed by inverting this forward model.
In the case of X-ray \gls{ct}, an image of the attenuation coefficient is reconstructed from projections acquired in a circular acquisition.
Under orthographic projection, this model becomes the Radon transform.
Discretized measurements of the Radon transform are called sinograms and their inversion is well-studied.
The most prominent approach is a direct inversion with a closed-form solution called \gls{fbp}~\cite{maier2018medical}.

\begin{figure}[t]
\begin{center}
   \includegraphics[width=\linewidth]{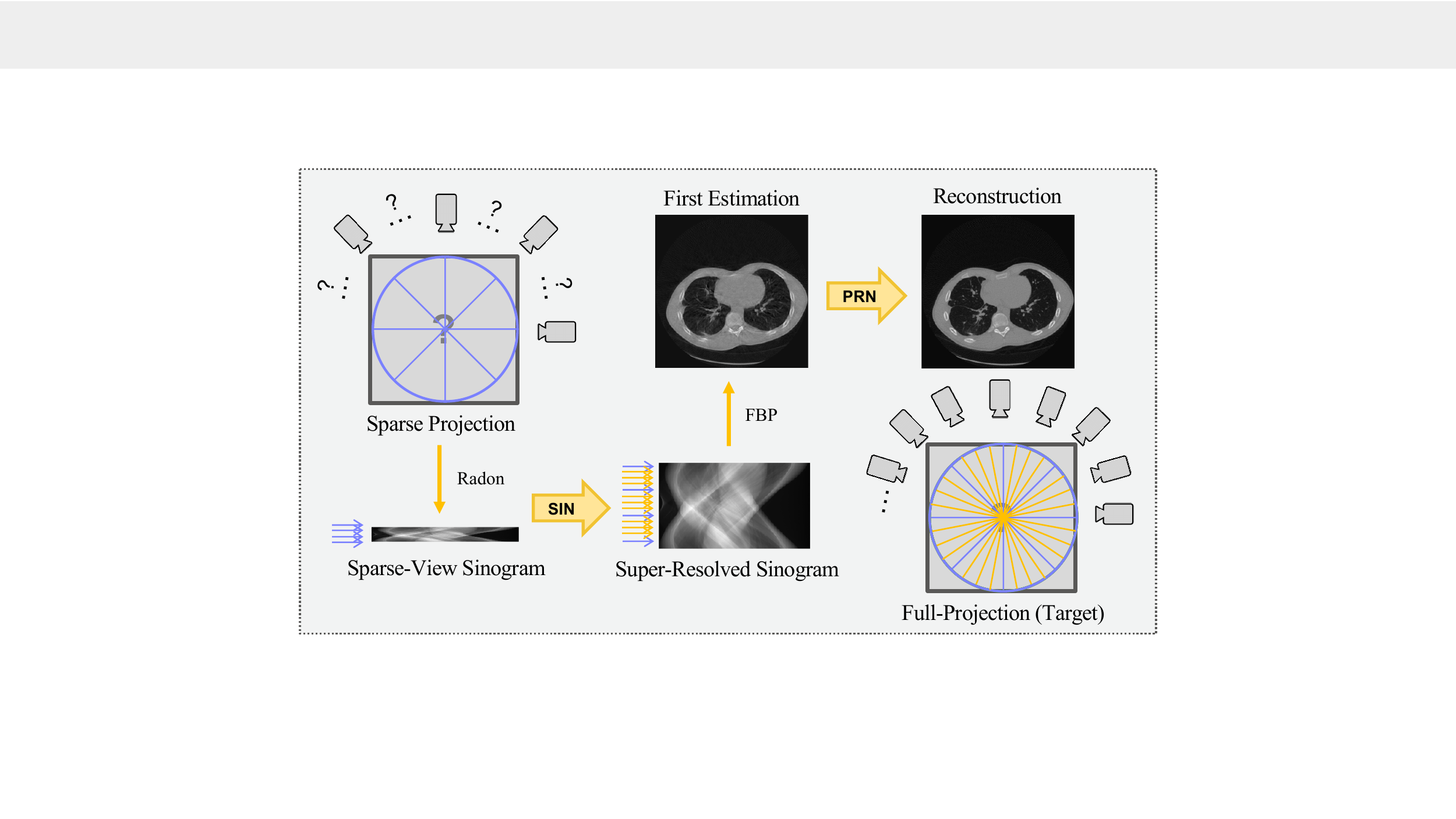}
\end{center}
   \caption{Proposed two-step strategy for the sparse-view CT problem, where projections are not enough to recover a high-quality reconstruction. To obtain a reliable reconstruction from sparse projections, a \gls{sin} first super-resolves the projection image in one dimension. This allows a direct reconstruction via~\gls{fbp}.
   Then the \gls{prn} removes remaining artifacts.}
\label{fig:abstract}
\end{figure}

Accurate reconstruction with \gls{fbp} requires sufficient angular sampling in terms of the Crowther-criterion~\cite{crowther1970reconstruction} derived from the Shannon-Nyquist sampling theorem.
If violated, aliasing artifacts appear as streak-artifacts that drastically reduce diagnostic quality.
This setting is called \gls{svct} and arises in many practical applications, addressing motion compensation~\cite{enjilela2019cubic} or dose-reduction~\cite{gao2014low,yu2009radiation}.

A number of approaches have been proposed to overcome the ill-posedness of \gls{svct} by incorporating prior knowledge into the reconstruction algorithm.
Traditionally, \gls{ir} is used to optimize a data-fidelity term coupled with appropriate priors~\cite{kudo2013image,zhu2013improved,vandeghinste2011split}.
For example, the popular \gls{tv} prior assumes piece-wise constant objects for \gls{ct}~\cite{sidky2008image}.
Recently, learning-based reconstruction algorithms have attracted a lot of attention~\cite{adler2017solving}.
Prominent examples are \gls{cnn}-based models designed to remove artifacts in reconstruction domain~\cite{kofler2018u} or unrolled networks mimicking the data-fidelity driven \gls{ir}~\cite{ongie2020deep}.

\begin{figure}[t]
\begin{center}
  \includegraphics[width=\linewidth]{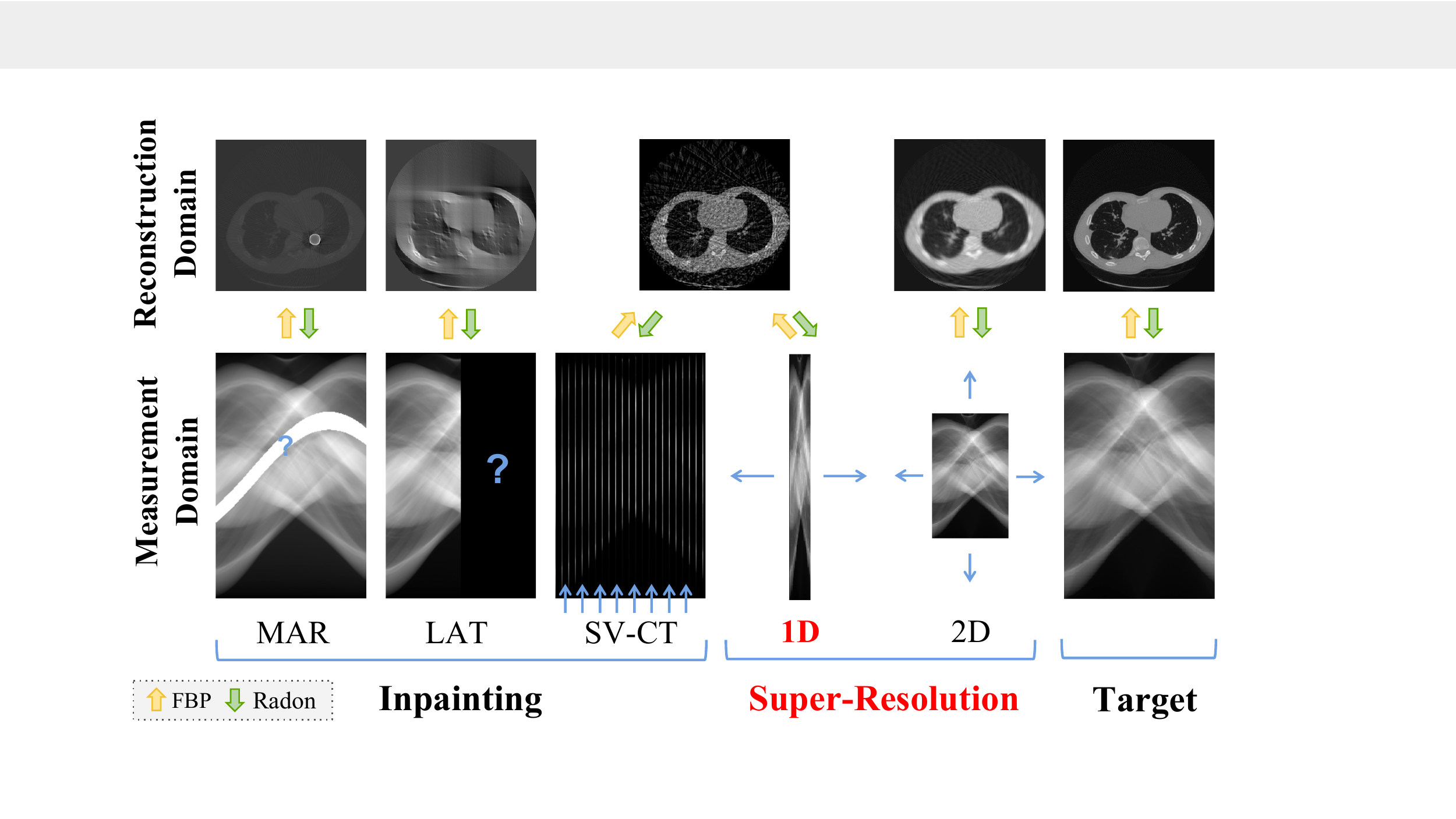}
\end{center}
  \caption{Different problems arising in tomographic imaging and their relations to our 1D-super-resolution task, such as metal artifact removal (MAR) and limited angle tomography (LAT). \gls{svct} is at the intersection of inpainting and super-resolution when viewed in measurement domain.}
\label{fig:superres_vs_inpainting}
\end{figure}

A less common approach is to interpret \gls{svct} as a super-resolution or inpainting problem, where the missing information is interpolated between the sparse projections. 
A sparse-view sinogram can be either inpainted or 1D-super-resolved.
%
Either way, 1D-upsampling of a sinogram is more constrained than its 2D-counterpart~\cite{katsaggelos2007super} of natural images, since the sinogram data vary smoothly along the angular dimension~\cite{helgason}.
%
Other typical tomographic problems with similar formulations include metal artifact removal and limited angle tomography, see Fig.~\ref{fig:superres_vs_inpainting}.
%
%

Sinogram inpainting did not attract much attention compared to iterative approaches.
This is because capturing the intricate representations of sinograms turns out to be an extremely challenging task~\cite{li2012strategy,tovey2019directional,kostler2006adaptive}.
%
However, the fast development of deep learning now enables us to efficiently capture low-dimensional representations of data in various image domains.
These technological advances motivate us to consider the possibilities of tackling \gls{svct} with learning-based super-resolution.
%

In particular, we argue and will demonstrate that image-processing in the measurement domain is favored over post-reconstruction processing.
%
This is because sinogram-based methods prevent the generation of streak artifacts that are difficult to remove using local processing (e.g., CNNs) because they are widely distributed over the image domain.

In this work, we propose a two-step architecture that learns super-resolution in the sinogram domain and then removes the remaining artifacts in the image domain.
The core idea is outlined in Fig.~\ref{fig:abstract}.
%
The following summarizes our main contributions:

\begin{itemize}
    \item A novel 2-step reconstruction approach in both measurement and image domain. It ensures maximum detail-preservation compared to single-step networks.
    \item A novel domain-specific, light-weight perceptual loss that outperforms the VGG-based perceptual loss~\cite{johnson2016perceptual} in terms of accuracy, memory and time efficiency. 
    \item Loss functions and submodules tailored to sinogram-domain tasks, and an extensive ablation study that analyzes the effectiveness of each proposed module.
    \item Demonstration of accurate reconstructions for high compression ratios while achieving a significant performance gain (over $4$ dB PSNR) compared to state-of-the-art algorithms.
\end{itemize}

\section{Related Work}

\subsubsection*{Learned Image Reconstruction}
Recent advances in the reconstruction community seek to learn better image priors from available data~\cite{ongie2020deep}.
%
Two prominent strategies stand out: \Gls{pn} and \Gls{obl}.
%

\begin{figure*}[t]
\begin{center}
    \includegraphics[width=\linewidth]{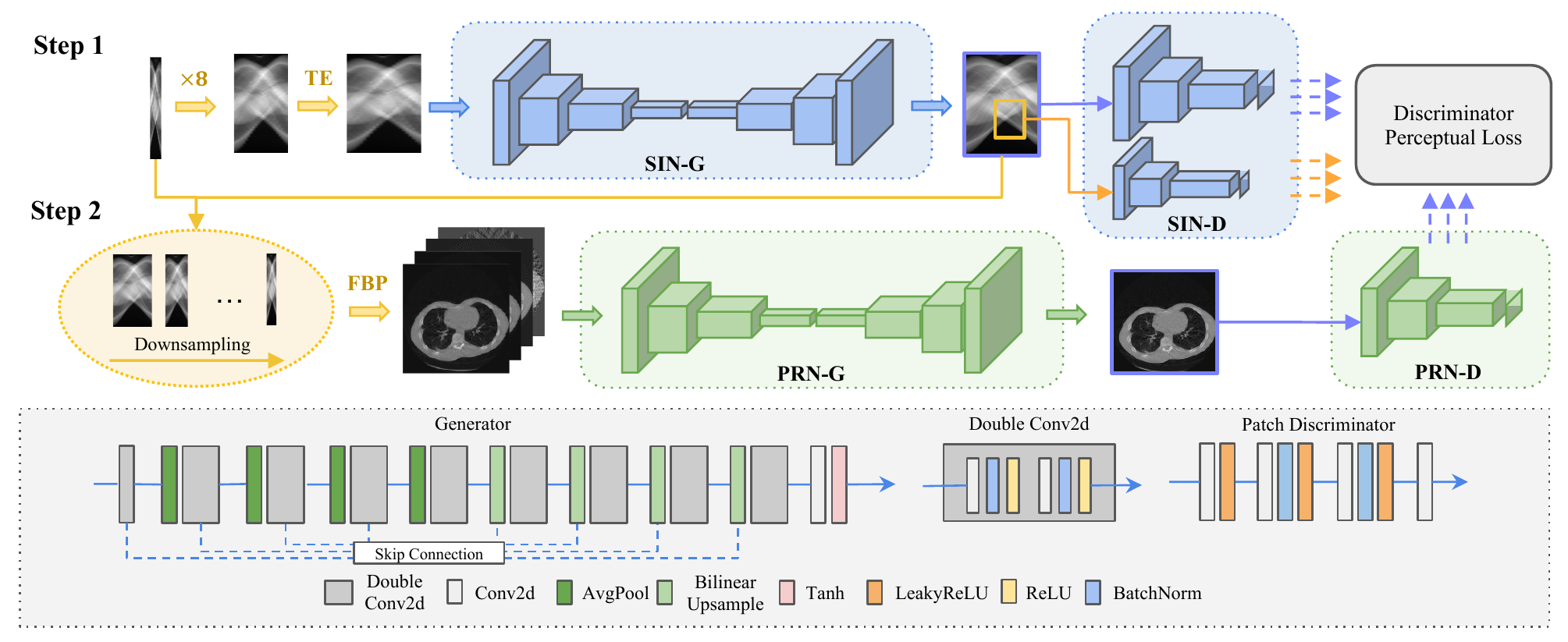}
\end{center}
   \caption{Network Architecture of our proposed SIN-4c-PRN model. \textbf{Top:} In step one, we linearly upsample a sparse-view sinogram and preprocess with our Two-Ends (TE) flipping method as input to the \gls{sin} model. In step two, the super-resolved sinogram is 1D-downsampled by different factors. \gls{fbp} reconstructions from the set of sinograms are concatenated as a cascaded input to the \gls{prn}.
   The patch discriminators SIN-D and PRN-D calculate pixel-wise adversarial losses of global or local generated images.
   Additionally, each discriminator calculates a discriminator perceptual loss that enhances the generator's perceptual knowledge. \textbf{Bottom:} Detailed architecture of our SIN and PRN models.}
\label{fig:model}
\end{figure*}

A single-step \gls{pn} removes aliasing artifacts after a direct inversion with closed-form solutions such as \gls{fbp}.
%
%
Variants of different neural networks are proposed to learn \gls{pn} models to eliminate streak artifacts, such as GoogLeNet~\cite{xie2018artifact}, DenseNet~\cite{8331861} or TomoGAN~\cite{liu2020tomogan}.
%
An improved version was presented by Han and Ye~\cite{han2018framing}, showing that networks fulfilling the frame-condition yield improved outcomes.
A downside of these \gls{pn}-based methods is their inability to constrain their output by relating it to the measured data.

With \gls{obl} a differentiable forward model is included into the network.
Most \gls{obl} frameworks unroll \gls{ir}, such as gradient descent, and learn the gradient of an image prior that is applied during each iteration.
For \gls{svct}, Chen \emph{et~al.}~\cite{chen2020airnet} demonstrated state-of-the-art performance with \gls{obl}.
A more general approach to learning a data-driven inversion of linear inverse problems are Neumann networks introduced by Gilton \emph{et~al.}~\cite{gilton2019neumann}.

An entirely different approach from iterative reconstruction with learned priors is to reuse an arbitrary denoising method for images as a proximal operator.
This approach was pioneered by Venkatakrishnan~\emph{et~al.}~\cite{venkatakrishnan2013plug}, referred to \gls{pp}-priors and can be combined with learned denoising.

\subsubsection*{Image Super-Resolution $\&$ Inpainting}

A different approach to solve \gls{svct} poses it as a super-resolution or inpainting problem in the measurement domain.
%
This is advantageous because sinograms do not yet suffer from aliasing artifacts.
%
U-Nets~\cite{ronneberger2015u} and \gls{gan}~\cite{goodfellow2014generative} have demonstrated strong performance in this task by either considering the whole sinogram~\cite{dong2019sinogram, ghani2018deep, tan2019sharpness,yoo2019sinogram} or local patches~\cite{lee2018deep}.

Recent approaches in other tomographic applications (Fig.~\ref{fig:superres_vs_inpainting}) combine learning in both sinogram and image domain to leverage the advantages of both representations.
For example, Zhao~\emph{et~al.}~\cite{zhao2018unsupervised} proposed an unsupervised sinogram inpainting network trained in both domains in a cycle for limited angle tomography.
%
One of the first to propose adding a Radon inversion layer to directly couple measurement and reconstruction domain was W{\'u}rfl~\emph{et~al.}\cite{wurfl2016deep}.  %
The idea of dual-domain training was borrowed by DuDoNet~\cite{lin2019dudonet} to first inpaint missing information due to metal-artifacts, see Fig.~\ref{fig:superres_vs_inpainting} and refine the subsequent reconstruction.

\subsubsection*{Perceptual Losses}

\gls{gan} loss functions are often paired with a perceptual loss to produce visually convincing images.
Perceptual loss was introduced in the seminal work by Johnson~\emph{et~al.}~\cite{johnson2016perceptual} to tackle the long-standing problem that pixel-based loss-functions, such as the one using the $l_2$-norm, struggling to capture higher-level feature information. 
Perceptual losses are increasingly incorporated in many imaging algorithms showing superior performance while maintaining a reasonable convergence speed~\cite{zhang2018unreasonable}.
Yet, there is an ongoing debate about whether the natural color image features apply to images with a drastically different formation model and diverse content.
In the medical imaging setting, domain-specific perceptual networks were proposed to account for the domain change~\cite{li2020sacnn, ouyang2019ultra}.
These methods require pre-training of an additional network which increases the complexity of the complete training pipeline, making it harder to reproduce results.

\section{Proposed Approach}

We propose a two-step framework with novel loss-functions tailored for \gls{svct}, shown in Fig.~\ref{fig:model}.
First, we augment a sparse-view sinogram by 1D linear-interpolation and Two-Ends preprocessing, which accounts for boundary artifacts. 
The \glsfirst{sin} generates a reliable super-resolved sinogram 
so that the object can be reconstructed without strong streak artifacts via \gls{fbp}.
While \gls{sin} successfully restores a full-view sinogram, small imperfections still lead to some less-severe, more localized artifacts.
In the final refinement step, our \glsfirst{prn} removes such artifacts to obtain a high-quality reconstruction. 

This section discusses the network architectures, introduces our novel discriminator-perceptual loss, and outlines the optimization procedure.

\subsection{Two-Ends Sinogram Flipping}

\begin{figure}[t]
\begin{center}
   \includegraphics[width=.9\linewidth]{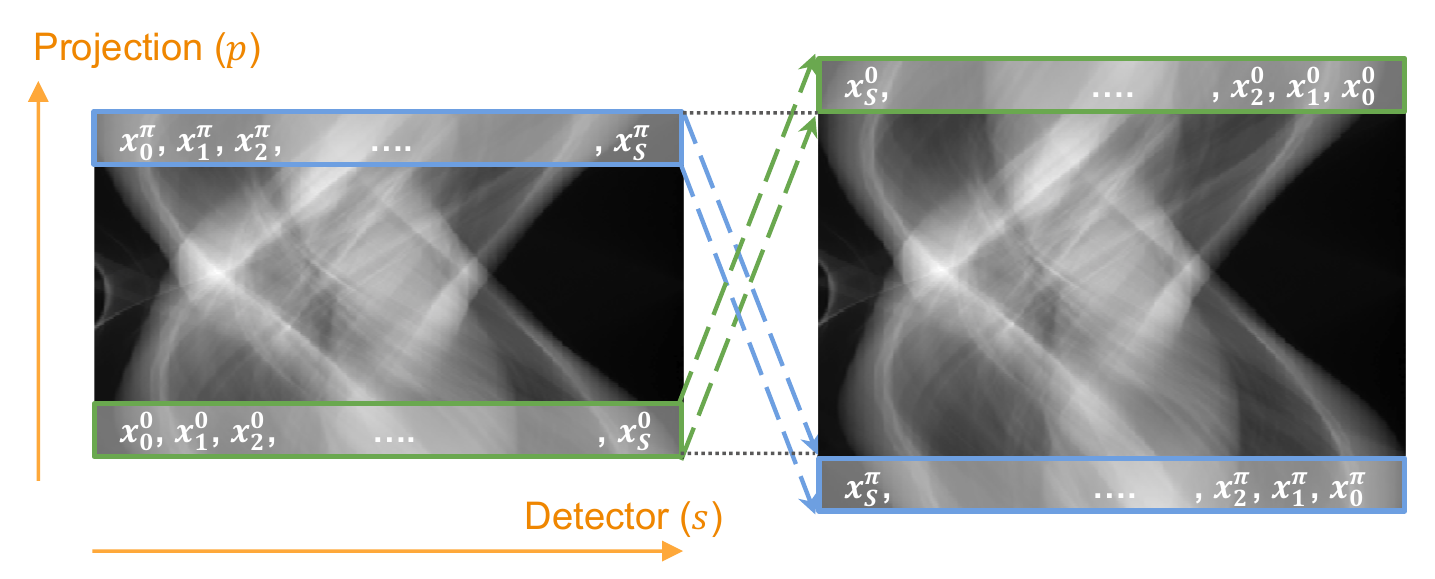}
\end{center}
   \caption{We avoid boundary artifacts during convolutions by reusing projections on both ends of the sinogram.}
\label{fig:two_ends}
\end{figure}

Boundary pixels have only one-sided neighborhood information. This quickly results in border blurriness detrimental for reconstruction.
We present a simple two-ends method that appends angles at the two ends of sparse-view sinograms as outlined in Fig.~\ref{fig:two_ends}.

Note that we only leverage data already present in the measurements. This is because under orthographic projection, the total intensity detected from angle $\theta$ is a flipped version of the projection at angle $\theta+\pi$.
This allows us to repeat the first and last few angles of a sinogram by flipping them along the detector axis and appending them to the other end. 

\subsection{Network Architectures}
\label{section:Network}

Following previous works in learned CT-reconstruction, we use the \gls{gan}~\cite{goodfellow2014generative} framework in both steps.
We first describe our adapted \mbox{U-Net} generators and global and local patch discriminators~\cite{iizuka2017globally, isola2017image} in detail.
Then we introduce our \gls{sin}-4c-\gls{prn} architecture based on \gls{sin} and \gls{prn}.

\subsubsection*{Adapted U-Net Generators}

A U-Net is a CNN that includes an encoding-decoding process with skip connections~\cite{ronneberger2015u}.
We made the following adaptions to the \mbox{U-Net} modules in our framework:
Four average-pooling and bilinear-upsampling layers, each followed by several deep convolutional layers. We keep skip-connections at each resolution level to ensure low-level feature consistency.

We find that resizing operations such as pooling and bilinear layers restore smooth sinusoids essential for tomographic reconstruction.
In contrast, half-strided convolutions introduce severe checkerboard artifacts~\cite{odena2016deconvolution} as in our experiments.

We further replace max-pooling with average-pooling, since a differentiable operator is preferable over subdifferentials for restoration tasks~\cite{zhou2016learning, yu2014mixed}.
The preservation of gradients maximally recovers measurement-domain information. Otherwise, if the high-frequencies produced by max-pooling are misaligned, errors will be amplified by the following FBP operation.

\subsubsection*{Patch Discriminators}
Inspired by Isola~\emph{et~al.}~\cite{isola2017image}, our discriminators include four convolution layers that downscale the input by a factor of four and output image patches.
Pixel-wise probabilities of these patches being real or fake are then calculated as adversarial losses described in Sec.~\ref{section:Optimization}.
We also keep the discriminators in relatively low dimensional feature space ($\leq256$ kernels) for ease of training without compromising performance~\cite{arjovsky2017towards}.

\subsubsection*{Sinogram Inpainting Network (SIN)}
The proposed \gls{sin} fills in the missing projections by directly learning from sparse-view to full-view sinograms.
The adapted U-Net generates super-resolved sinograms from an initial linear estimation, trained simultaneously with two patch discriminators that focus on either global or local descriptions of the generated sinograms.
Specifically, the local discriminator randomly picks $1/4 \times 1/4$ local patches from super-resolved sinograms to enhance local sinusoidal details.

\subsubsection*{Postprocessing Refinement Network (PRN)}

\begin{figure}[t]
\centering
   \includegraphics[width=\linewidth]{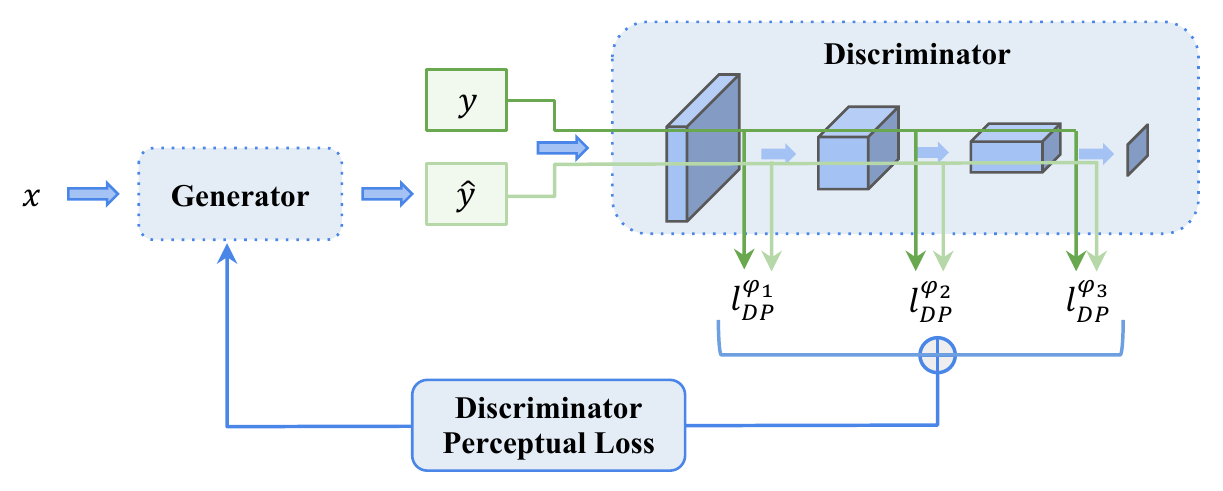}
   \caption{Discriminator Perceptual Loss. The image pair $\hat y$ generated from the input $x$, and the target image $y$ are compared at each activation layer output $\phi_j$ in the discriminator. Pixel-wise losses are calculated for updating the generator. }
\label{fig:perceptual}
\end{figure}

\gls{prn} is connected to \gls{sin} by the \gls{fbp} operator,
and it includes a U-Net and a patch discriminator.

One major consideration in designing a second-step network is to prevent propagation of errors introduced by our \gls{sin}. To solve this, we introduce a set of reconstructions from multi-resolution sinograms for our PRN to learn from.
We observe that \gls{svct} reconstructions already show sharp partial information although severely obscured by aliasing artifacts. Effective use of such information can be achieved by constructing a pool of diverse features for the network to learn from. 

In particular, we downsample the \gls{sin}-inpainted sinograms along the angular axis by factors of 2 and~4.
Then four different \gls{fbp} reconstructions from sparse, $2\times$downsampled, $4\times$downsampled, and fully-inpainted sinograms are concatenated as a 4-channel input to the \gls{prn}.

\subsection{Discriminator Perceptual Network}
\label{section:DP}

Perceptual losses have proven to significantly enhance the ability of networks to generate images with high perceptual quality~\cite{zhang2018unreasonable} when compared to traditional pixel-wise losses.
The original perceptual loss computes the $l_2$ norm of generated and target image pairs through a pre-trained \gls{vgg} network~\cite{johnson2016perceptual}, which serves as a feature extractor that enforces high-level feature fidelity.

We propose a novel \gls{dp} loss that is easy to incorporate into the formulation and especially suitable for domain-specific tasks.
It interprets the first few layers of a discriminator as a feature extractor, which is trained simultaneously with the generator on the current problem, and encourages feature-level similarity, see Fig.~\ref{fig:perceptual}.
The \gls{dp}-loss is computed in a similar fashion as the original perceptual-loss with \gls{vgg}.

We find out that introducing the \gls{dp}-loss promotes stability in our GAN training procedure.
For example, we can train the discriminator until convergence at each iteration of the minimax game.
An incremental strategy for training GANs with \gls{dp}-loss is provided in the supplementary material for further reference.

DP loss is calculated by the $l_2$ norms of the error between outputs of each activation layer $\phi_j$ in the discriminator with respect to the input pair consisting of the generated image $\hat y$ and the target image $y$. While one can choose any such layer $j$ in a discriminator, we compute the averaged norm for each $\phi_j$ before the last one (i.e., we do not include the discriminator output layer):
\begin{equation}
\mathcal{L}_{DP}(\hat y, y) = \frac{1}{N}\sum_{j=1}^N\frac{1}{C_jH_jW_j} \| \phi_j(\hat y) - \phi_j(y)\|_2^2
\end{equation}
where $C_j$ is the number of channels, $H_j$, $W_j$ are height and width of the output at $\phi_j$, respectively, and $N$ is the number of activation layers we use.

\subsection{Optimization}
\label{section:Optimization}

\subsubsection*{Adversarial loss} 

In the GAN-framework~\cite{goodfellow2014generative}, the generator is augmented by a discriminator that discerns real and synthesized images.
During training, the generator and the discriminator compete in a min-max game that leads to the adversarial loss.

Given a corrupted image $x$ and a target image $y$, we formulate the optimization target as follows:
\begin{equation}
\begin{aligned}
\min _G \max _{D} V(G,D) = &\mathbb{E}_{y \sim target} [\log D(y)] + \\
    &\mathbb{E}_{x \sim corrupt} [\log (1-D(G(x))],
\end{aligned}
\label{equation:GAN}
\end{equation}
where $G$ and $D$ represent the generator and discriminator networks, respectively.

Specifically, in \gls{sin}, the input $x \in \mathbb{R}^{S \times P}$ is the preprocessed sparse-view sinogram to $S$ detector pixels and $P$ angles and target $y \in \mathbb{R}^{S \times P'}$ is a preprocessed full-view sinogram.
Local discriminators compare $G(x)_{patch}$ and $y_{patch} \in \mathbb{R}^{\floor{S/4} \times \floor{P'/4}}$, which are small patches from the sinogram randomly chosen from $G(x)$ and $y$, respectively.
In PRN, $x$ and $y \in \mathbb{R}^{S \times S}$ form the input and target reconstructions pair.

\subsubsection*{Content Loss}

Besides the adversarial loss, we ensure the data fidelity of the generated images by applying an $l_1$-norm based pixel-reconstruction loss:
\begin{equation}
\begin{split}
\mathcal{L}_c(\hat y, y) = \|\hat y - y\|_1 ,
\end{split}
\label{equation:content}
\end{equation}
where $ \hat y $ is the generated and $y$ the target image.
The content loss is calculated in both \gls{sin} and \gls{prn}.
As suggested in~\cite{zhao2016loss}, we choose the $l_1$ loss since our experiments showed slightly sharper reconstructions compared to the $l_2$ loss.

\subsubsection*{High-Frequency (HF) Loss}

\begin{figure}[t]
\begin{center}
   \includegraphics[width=.9\linewidth]{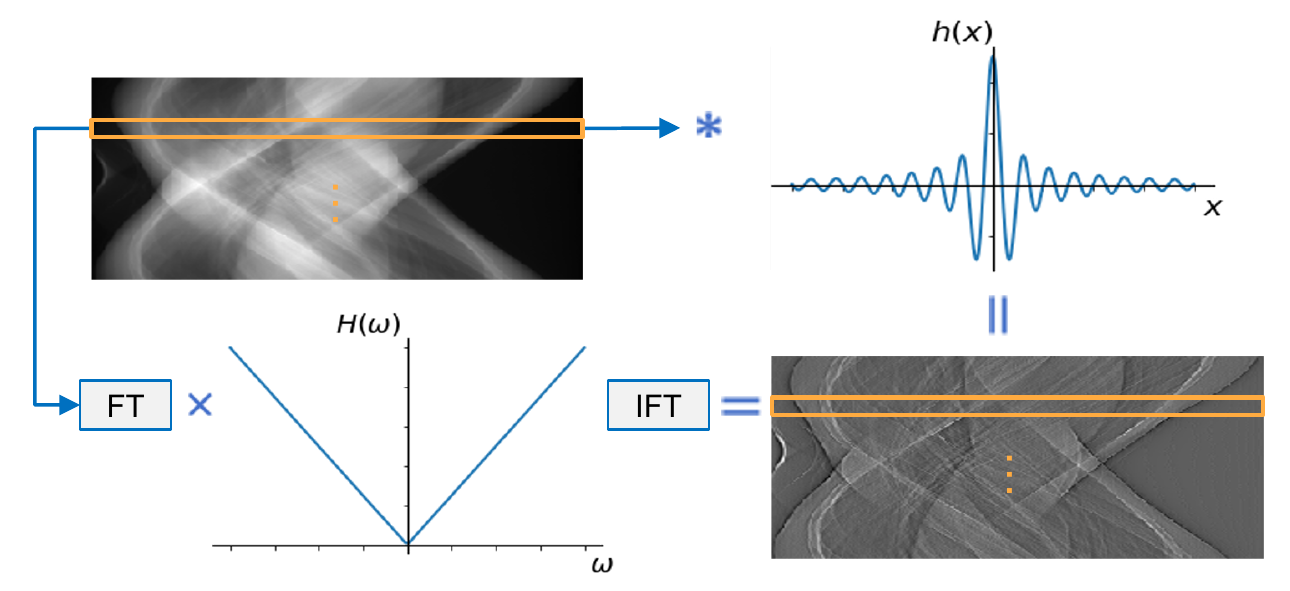}
\end{center}
   \caption{Applying Ramp filter to a sinogram preserves high frequency features.
   The filter is applied along one dimension in the frequency or spatial domains.
   Since the Ramp operator is back-propagatable~\cite{torch_radon}, losses can be defined on the ramp-filtered signal.}
\label{fig:ramp}
\end{figure}

Neural networks tend to prioritize learning of low-frequency features during training~\cite{rahaman2019spectral}.
While it might be acceptable for \gls{prn} since tomographic reconstructions tend to be relatively smooth, it is of critical importance for \gls{sin} where high-frequency information is essential for correct reconstructions.
This is because \gls{fbp} requires ramp filtering~\cite{maier2018medical} which is a high-pass filter compensating for oversampling the low-frequencies, as illustrated in Fig.~\ref{fig:ramp}.

To incorporate this CT-domain knowledge, we enforce our \gls{sin} to restore an accurate ramp-filtered sinogram by defining a high-frequency loss:
\begin{equation}
    \mathcal{L}_{HF}(\hat y, y) = \|\hat y \ast h  - y \ast h\|_1,
\end{equation}
where $h$ is the ramp kernel in the spatial domain, and $\ast$ denotes convolution.
Note that the high-frequency loss is fully backpropagatable~\cite{torch_radon}.

\subsubsection*{Total Objective}

Finally, the total objective function of the \gls{sin} generator is the weighted sum of the four losses mentioned above:

\begin{equation}
    G_{SIN}\text{*} = arg \min _G [\lambda_1 \mathcal{L}_{adv}(G) + \lambda_2 \mathcal{L}_c + \lambda_3 \mathcal{L}_{DP} + \lambda_4 \mathcal{L}_{HF}],
\label{equation:total_SIN}
\end{equation}
where $\mathcal{L}_{adv}$ defines the adversarial loss~\cite{goodfellow2014generative} (details in the supplementary material) and $\lambda_i$ are empirically chosen weighting parameters.
Based on our experiments, we set $\lambda_1 = 1, \lambda_2 = 50, \lambda_3 = 20, \lambda_4 = 50$. All discriminator adversarial losses are assumed a weight equal to one.

Similarly, the total objective function for the \gls{prn} is:
\begin{equation}
\begin{aligned}
    G_{PRN}\text{*} &= arg\min_G [\lambda_1 \mathcal{L}_{adv}(G) + \lambda_2 \mathcal{L}_c + \lambda_3 \mathcal{L}_{DP}],
\label{equation:total_PRN}
\end{aligned}
\end{equation}
where all parameters $\lambda_i$ are chosen the same as for \gls{sin}.

\section{Experiments}
\label{section:experiments}

Our experiments are performed on the open-source TCIA LDCT-and-Projection-Dataset~\cite{clark2013cancer, mccollough2020low}.
We extensively evaluate our methodology both qualitatively using samples of chest, abdomen, and head images and quantitatively using the \gls{ssim} and \gls{psnr}.

We define a full-view target sinogram $y^* \in \mathbb{R}^{S \times P^*}$ to have $S = 320$ detector pixels and $P^* = 180$ uniformly-distributed angles over $\pi$.
The sparse-view input includes 23 angles sampled from $y^*$ once every eight projections,
and the reconstructed image have a resolution of \(320 \times 320\) pixels.
Due to the proposed two-ends preprocessing, we take six angles from each end of $y^*$ to generate a 192-angle sinogram $y \in \mathbb{R}^{S \times (P^*+12)} $ as the ground-truth for \gls{sin}.

\subsection{Datasets}

Our dataset contains 5394 images resized to \(320 \times 320\) stemming from 68 patients taken from the TCIA dataset.
Additionally, we augment each sample with a random affine transform to increase data diversity during training.
Note that this transformation is performed on the reconstructions, which leads to significant variation in the sinogram domain.
We then extract a set of around 500 images from the dataset patient-wise for testing.
Once we choose a patient for testing, we remove all data related to this patient from the training set to avoid inter-slice correlation.
Full-view sinograms are generated using parallel-beam (orthographic) projections over an angular range of 180 degree, which define the target set for \gls{sin}.

\subsection{Training Details}
\label{section: Training Details}

Following the standard \gls{gan} procedure \cite{goodfellow2014generative,iizuka2017globally}, the generator and both global and local discriminators are trained alternatively.
The ADAM~\cite{kingma2014adam} solver with momentum parameters \(\beta_1 = 0.5\) and \(\beta_2 = 0.999\) are used to optimize all networks.
We set the learning rate to 0.0001 and run 100 epochs or until convergence for all experiments.
For further details on the training procedure, please consult the supplementary material.

\begin{figure}
    \centering
    \begin{subfigure}[t]{.32\linewidth}
        \centering
        \begin{overpic}[width=\linewidth]{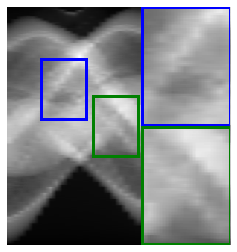}
         \put (17,6) {\textcolor{orange}{33.56}}
         \put (17,16) {\textcolor{orange}{0.917}}
        \end{overpic}
        \caption{Linear}
    \end{subfigure}
    \begin{subfigure}[t]{.32\linewidth}
        \centering
        \begin{overpic}[width=\linewidth]{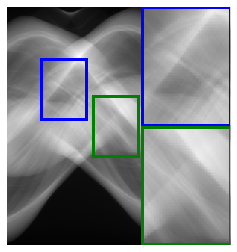}
         \put (17,6) {\textcolor{orange}{40.22}}
         \put (17,16) {\textcolor{orange}{0.978}}
        \end{overpic}
        \caption{SIN}
    \end{subfigure}
    \begin{subfigure}[t]{.32\linewidth}
        \centering
        \begin{overpic}[width=\linewidth]{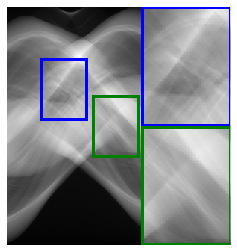}
         \put (17,6) {\textcolor{orange}{PSNR}}
         \put (17,16) {\textcolor{orange}{SSIM}}
        \end{overpic}
        \caption{Ground Truth}
    \end{subfigure}
    \begin{subfigure}[t]{.32\linewidth}
        \centering
        \begin{overpic}[width=\linewidth]{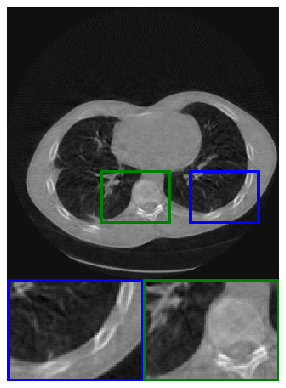}
         \put (5,90) {\textcolor{orange}{31.96}}
         \put (48,90) {\textcolor{orange}{0.828}}
        \end{overpic}
        \caption{SIN}
        \label{fig:results_SIN_FBP}
    \end{subfigure}
        \begin{subfigure}[t]{.32\linewidth}
        \centering
        \begin{overpic}[width=\linewidth]{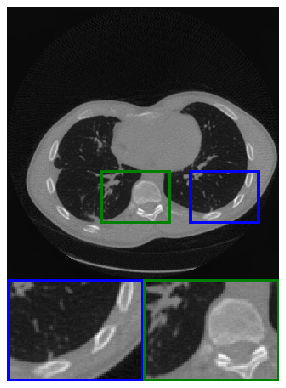}
         \put (5,90) {\textcolor{orange}{33.15}}
         \put (48,90) {\textcolor{orange}{0.859}}
        \end{overpic}
        \caption{SIN-4c-PRN}
    \end{subfigure}
    \begin{subfigure}[t]{.32\linewidth}
        \centering
        \begin{overpic}[width=\linewidth]{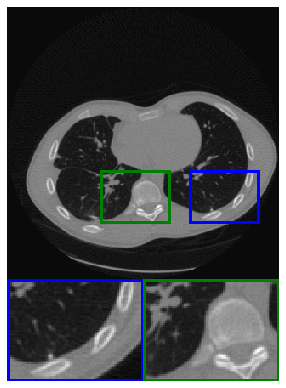}
         \put (5,90) {\textcolor{orange}{PSNR}}
         \put (48,90) {\textcolor{orange}{SSIM}}
        \end{overpic}
        \caption{Ground Truth}
    \end{subfigure}
    \caption{Our results in both measurement and reconstruction domains with zoom-in details compared to linear interpolation.}
    \label{fig:results}
\end{figure}

\begin{figure*}[t]
\centering
    \begin{subfigure}[t]{.162\linewidth}
\centering
\begin{overpic}[width=\linewidth]{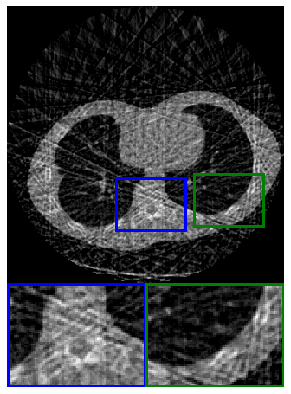}
 \put (5,90) {\textcolor{orange}{21.52}} 
 \put (50,90) {\textcolor{orange}{0.438}} 
\end{overpic}
    \end{subfigure}
    \begin{subfigure}[t]{.162\linewidth}
        \centering
\begin{overpic}[width=\linewidth]{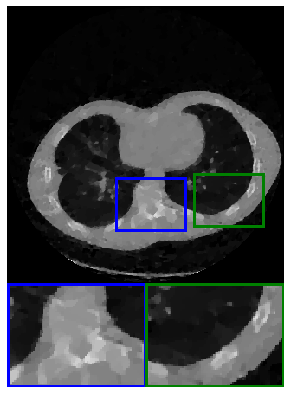}
 \put (5,90) {\textcolor{orange}{29.90}}
 \put (50,90) {\textcolor{orange}{0.805}}
\end{overpic}
    \end{subfigure}
    \begin{subfigure}[t]{.162\linewidth}
        \centering
\begin{overpic}[width=\linewidth]{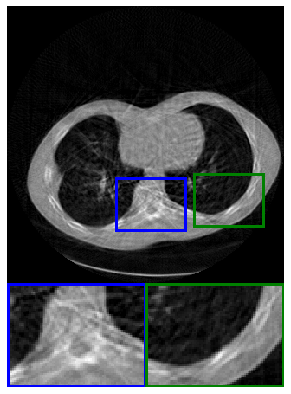}
 \put (5,90) {\textcolor{orange}{28.87}}
 \put (50,90) {\textcolor{orange}{0.731}}
\end{overpic}
    \end{subfigure}
    \begin{subfigure}[t]{.162\linewidth}
        \centering
\begin{overpic}[width=\linewidth]{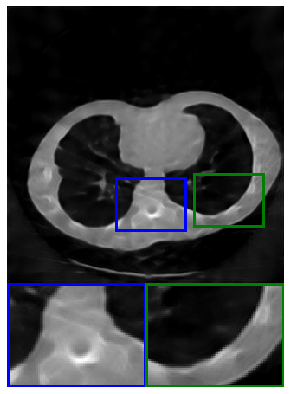}
 \put (5,90) {\textcolor{orange}{27.35}}
 \put (50,90) {\textcolor{orange}{0.619}}
\end{overpic}
    \end{subfigure}
    \begin{subfigure}[t]{.162\linewidth}
        \centering
\begin{overpic}[width=\linewidth]{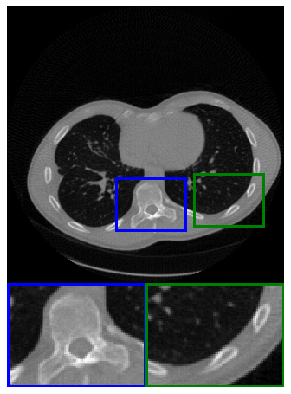}
 \put (5,90) {\textcolor{orange}{32.76}}
 \put (50,90) {\textcolor{orange}{0.861}}
\end{overpic}
    \end{subfigure}
    \begin{subfigure}[t]{.162\linewidth}
        \centering
\begin{overpic}[width=\linewidth]{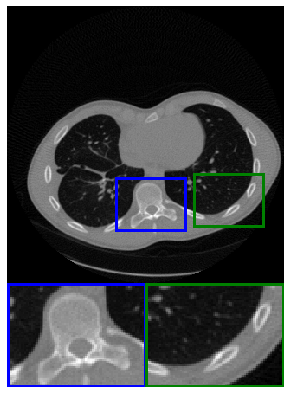}
 \put (5,90) {\textcolor{orange}{PSNR}}
 \put (50,90) {\textcolor{orange}{SSIM}}
\end{overpic}
    \end{subfigure}
    \begin{subfigure}[t]{.162\linewidth}
        \centering
\begin{overpic}[width=\linewidth]{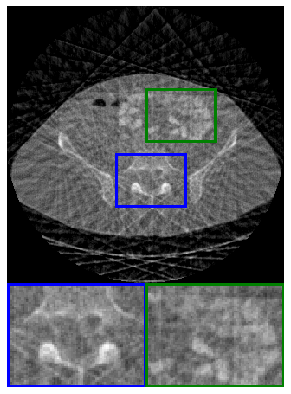}
 \put (5,90) {\textcolor{orange}{24.14}} 
 \put (50,90) {\textcolor{orange}{0.522}} 
\end{overpic}
    \end{subfigure}
    \begin{subfigure}[t]{.162\linewidth}
        \centering
\begin{overpic}[width=\linewidth]{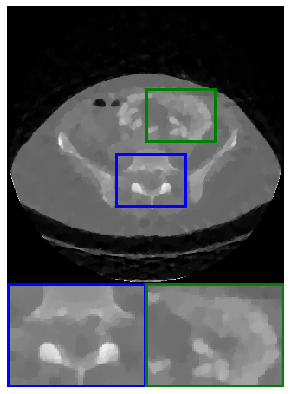}
 \put (5,90) {\textcolor{orange}{30.94}}
 \put (50,90) {\textcolor{orange}{0.847}}
\end{overpic}
    \end{subfigure}
    \begin{subfigure}[t]{.162\linewidth}
        \centering
\begin{overpic}[width=\linewidth]{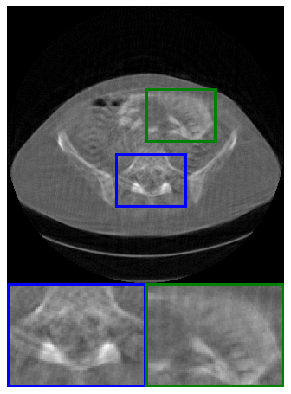}
 \put (5,90) {\textcolor{orange}{32.16}}
 \put (50,90) {\textcolor{orange}{0.796}}
\end{overpic}
    \end{subfigure}
    \begin{subfigure}[t]{.162\linewidth}
        \centering
\begin{overpic}[width=\linewidth]{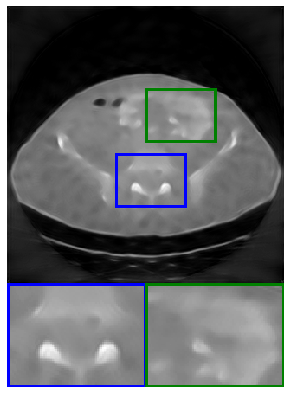}
 \put (5,90) {\textcolor{orange}{29.68}}
 \put (50,90) {\textcolor{orange}{0.788}}
\end{overpic}
    \end{subfigure}
    \begin{subfigure}[t]{.162\linewidth}
        \centering
\begin{overpic}[width=\linewidth]{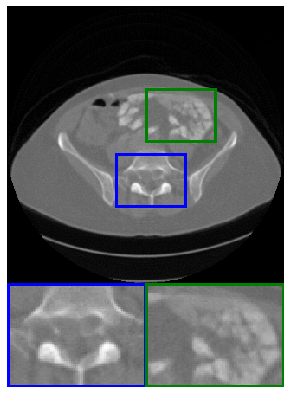}
 \put (5,90) {\textcolor{orange}{36.29}}
 \put (50,90) {\textcolor{orange}{0.913}}
\end{overpic}
    \end{subfigure}
    \begin{subfigure}[t]{.162\linewidth}
        \centering
\begin{overpic}[width=\linewidth]{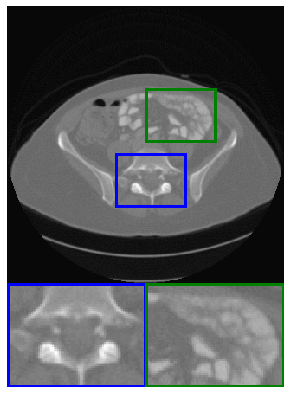}
 \put (5,90) {\textcolor{orange}{PSNR}}
 \put (50,90) {\textcolor{orange}{SSIM}}
\end{overpic}
    \end{subfigure}
    \begin{subfigure}[t]{.162\linewidth}
        \centering
\begin{overpic}[width=\linewidth]{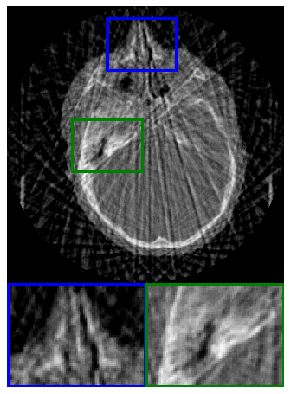}
 \put (5,90) {\textcolor{orange}{23.05}} 
 \put (50,90) {\textcolor{orange}{0.490}} 
\end{overpic}
        \caption{Sparse-View}
    \end{subfigure}
    \begin{subfigure}[t]{.162\linewidth}
        \centering
\begin{overpic}[width=\linewidth]{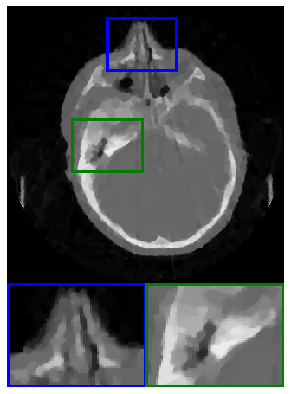}
 \put (5,90) {\textcolor{orange}{32.10}}
 \put (50,90) {\textcolor{orange}{0.867}}
\end{overpic}
        \caption{FISTA-TV~\cite{FISTA}}
    \end{subfigure}
    \begin{subfigure}[t]{.162\linewidth}
        \centering
\begin{overpic}[width=\linewidth]{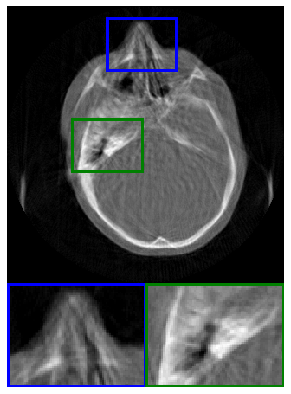}
 \put (5,90) {\textcolor{orange}{30.19}}
 \put (50,90) {\textcolor{orange}{0.778}}
\end{overpic}
        \caption{cGAN~\cite{ghani2018deep}}
    \end{subfigure}
    \begin{subfigure}[t]{.162\linewidth}
        \centering
\begin{overpic}[width=\linewidth]{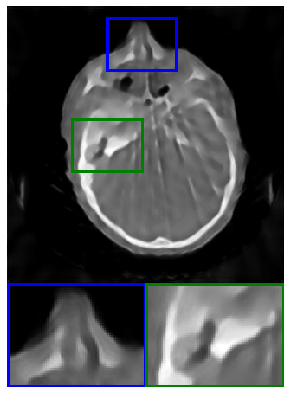}
 \put (5,90) {\textcolor{orange}{27.76}}
 \put (50,90) {\textcolor{orange}{0.650}}
\end{overpic}
        \caption{Neumann~\cite{gilton2019neumann}}
    \end{subfigure}
    \begin{subfigure}[t]{.162\linewidth}
        \centering
\begin{overpic}[width=\linewidth]{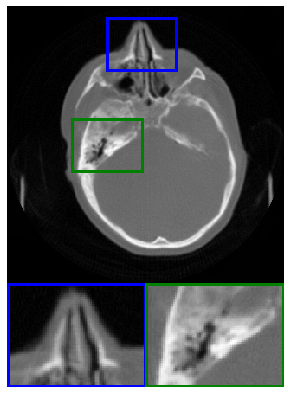}
 \put (5,90) {\textcolor{orange}{34.31}}
 \put (50,90) {\textcolor{orange}{0.915}}
\end{overpic}
        \caption{SIN-4c-PRN}
    \end{subfigure}
    \begin{subfigure}[t]{.162\linewidth}
        \centering
\begin{overpic}[width=\linewidth]{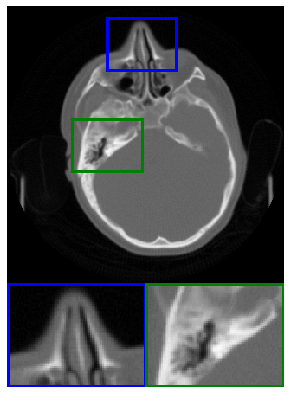}
 \put (5,90) {\textcolor{orange}{PSNR}}
 \put (50,90) {\textcolor{orange}{SSIM}}
\end{overpic}
        \caption{Ground Truth}
    \end{subfigure}
    
    \caption{Baseline results with zoom-in details. From top to bottom row: 
    \textbf{Chest:} In our SIN-4c-PRN result, both soft tissues in the black region of the body and sharp edges of bones are maximally recovered compared to other methods.
    \textbf{Abdomen:} Soft tissues with low contrast, such as in the green box are hard to recover from sinograms and pose a significant challenge for our networks.
    \textbf{Head:} The high-frequency details in the nose and bone regions are reconstructed while others either fail or over-smooth them.
    }
    \label{fig:baseline}
\end{figure*}
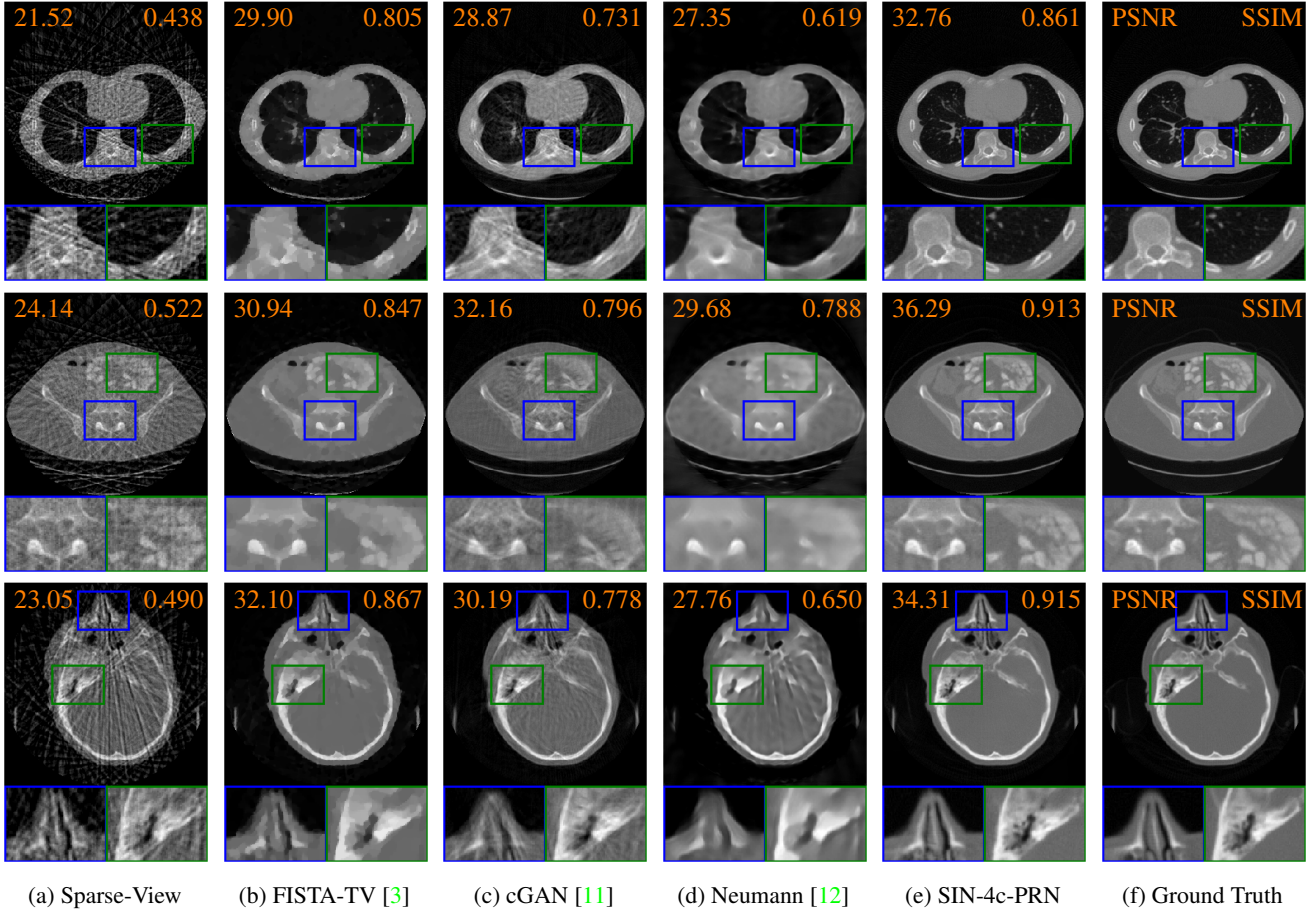

\subsection{Qualitative Results}

We first show our results in both measurement and reconstruction domains in Fig.~\ref{fig:results}.
Compared with a naive linear interpolation, our super-resolved sinograms accurately restore the smooth sinusoidal structures.
It is nearly impossible to discern difference between SIN-generated and ground-truth sinograms.
A subsequent reconstruction in Fig.~\ref{fig:results_SIN_FBP} obtained by \gls{fbp} shows an artifact-reduced reconstruction that resembles the target image.
However, there remains a novel type of less severe artifacts.
These localized artifacts can be efficiently removed by \gls{prn} in the second step.
The final result of our SIN-4c-PRN preserves details such as bones and soft-tissue contrast.

\subsubsection*{Comparison to State-of-the-Art}
In Fig.~\ref{fig:baseline}, we further compare our SIN-4c-PRN model to state-of-the-art CT reconstruction approaches.

FISTA-TV~\cite{FISTA} is a powerful iterative reconstruction algorithm that employs \gls{tv}-based regularization.
We use the implementation in the open-source toolkit ToMoBAR~\cite{kazantsev2020ctmeeting}.
To select the optimal hyperparameters of FISTA-TV, we perform a grid-search and choose the set with the best visual results.
FISTA-TV preserves edges well, but the reconstructions are over-smoothed, which is a well-known weakness of the \gls{tv}-prior.

We further compare with a learning-based sinogram inpainting network that uses a conditional-GAN (cGAN) based on an Encoding-Decoding U-Net~\cite{ghani2018deep} with strided-convolutions.
We implement their architecture and tuned hyper-parameters to obtain the best results.
Our experiments show that checkerboard artifacts still exist in generated sinograms due to the strided convolutions, see supplementary material for example images. And these artifacts are further amplified in reconstructions.
Note that the original work uses much lower undersampling factors than we are aiming for in this paper, which might explain the discrepancy to their findings.

Finally, we compare against our own PyTorch implementation of Neumann networks~\cite{gilton2019neumann}.
Neumann networks are a variant of unrolled optimization that learns a \gls{cnn}-based regularizer.
They have proven to yield great results in solving super-resolution and MRI problems. 
Note that unrolling networks require a back-propagatable tomography operator, for which we use torch-radon~\cite{torch_radon}.
Despite extensive tuning of hyper-parameters, our implementation fails to remove streak artifacts and over-smoothness.

\subsection{Quantitative Results}

For a quantitative evaluation we measure the \gls{psnr} and \gls{ssim} statistics within a circular region-of-interest using our test dataset.
Table~\ref{table:baselines} shows that our proposed SIN-4c-PRN model significantly improves the reconstruction accuracy compared to the other state-of-the-art reconstruction methods discussed earlier.
In particular, our model improves the average \gls{psnr} by over $4$ dB and \gls{ssim} by $5\%$, with narrowed variance.

\begin{table}[tb]
\centering
\begin{tabularx}{\linewidth}{|X|c|c|}
\hline
Method & PSNR ($\sigma$) & SSIM ($\sigma$) \\
\hline\hline
FISTA-PD-TV \cite{FISTA} & 30.61 (2.67) & 0.839 (0.036)\\
cGAN \cite{ghani2018deep} & 30.86 (1.92) & 0.762 (0.043)\\
Neumann Network \cite{gilton2019neumann} & 28.72 (2.09) & 0.697 (0.069)\\
Ours & \textbf{34.90 (2.15)} & \textbf{0.877 (0.029)}\\
\hline
\end{tabularx}
\caption{Performance of baseline models compared with our SIN-4c-PRN model.}
\label{table:baselines}
\end{table}

\section{Ablation Study}
\label{section:ablation}

This paper proposes several submodules which provide the best results when all used in conjunction.
Yet, the complete framework reaches a certain complexity and raises the question of whether all sub-modules actually improve performance.
For this reason, we carry out an extensive ablation study to analyze the impact of each module.
All of the experiments follow the training strategy in Sec. \ref{section: Training Details}, and are evaluated in the reconstruction domain.

\subsubsection*{Analysis among Models}

First, we compare the results of different architectures.
We raise the following questions about the necessity of our two-step networks: How good are the reconstructions if one only trains \gls{sin}, \gls{prn} (i.e., with a sparse-view reconstruction as input) or a SIN-PRN without the cascaded 4-channel input?
To validate, we train the four networks and summarize the quantitative results in Table~\ref{table:metric}.
We first show an improvement in both metrics of \gls{sin} compared to \gls{prn}, which is consistent with our arguments that measurement domain is easier to learn and preserves information. 
With both, \gls{sin}-\gls{prn} provides cleaner reconstructions.
The experiments further suggest that including the cascaded input improves both metrics and image quality.
We verify that the improvement of SIN-4c-PRN indeed comes from the diversity of inputs, by training another network with the same capacity but only single form of input.
For more training details and qualitative results, we refer to supplementary material.

\begin{table}
\centering
\begin{tabularx}{\linewidth}{|X|c|c|}
\hline
Model & PSNR ($\sigma$) & SSIM ($\sigma$) \\
\hline\hline
Sparse-View & 22.80 (1.77) & 0.485 (0.042)\\
PRN & 32.95 (1.96) & 0.835 (0.034)\\
SIN & 34.19 (2.35) & 0.859 (0.036)\\
SIN-PRN & 34.61 (2.14) & 0.873 (0.035)\\
SIN-4c-PRN & \textbf{34.90 (2.15)} & \textbf{0.877 (0.029)}\\
\hline
\end{tabularx}
\caption{Reconstruction performance with different composition of our models.}
\label{table:metric}
\end{table}

\subsubsection*{Analysis of Proposed Modules}

We perform controlled experiments to analyze the effectiveness of our two-ends (TE) preprocessing and high-frequency (HF) loss with a SIN model. Table \ref{table:metric_module} shows that with either module, \gls{psnr} improves by $2.5$ dB and \gls{ssim} by $5\%$.

\begin{table}
\centering
\begin{tabular}{|c|c|c|c|c|c|}
\hline
\multicolumn{2}{|c|}{Module} &\multicolumn{2}{|c|}{Perceptual} & \multicolumn{2}{|c|}{Metrics}\\
\hline\hline
TE & HF &DP &VGG & PSNR ($\sigma$) & SSIM ($\sigma$)   \\
\hline
\redmark     & \redmark     & \greencheck   &\redmark   & 30.84 (1.87) & 0.785 (0.046)\\
\redmark       & \greencheck &\greencheck & \redmark    & 31.65 (2.15) & 0.812 (0.050)\\
\greencheck  &  \redmark      &\greencheck & \redmark & 31.71 (2.14) & 0.808 (0.049)\\
\greencheck  & \greencheck &\redmark & \redmark    &33.39 (2.09) &0.829 (0.037) \\
\greencheck  & \greencheck & \redmark    &\greencheck &33.51 (2.02) &0.852 (0.036) \\
\greencheck  & \greencheck &\greencheck & \redmark    & \textbf{34.19 (2.35)} &\textbf{0.859 (0.036)} \\
\hline
\end{tabular}
\caption{Reconstruction performance of different sub-modules and perceptual losses trained on single SIN model.}
\label{table:metric_module}
\end{table}


\subsubsection*{Analysis of Discriminator Perceptual Loss}

We further compare our DP loss with VGG16 perceptual loss \cite{johnson2016perceptual} by simply substituting the perceptual loss modules in our SIN. Table~\ref{table:metric_module} includes comparison among non-perceptual loss, VGG16 and DP loss on sinogram data. Our DP network shows the best results on both metrics.
Qualitative results and sinogram domain metrics in the supplementary material show that the inpainted sinograms with DP are considerably better than with VGG16 and lose fewer details, especially in low-contrast regions.




\section{Conclusion}

In this paper, we propose a two-step reconstruction framework for sparse-view CT problems.
Our SR-network successfully learns the challenging task of upsampling a thin sinogram. 
The subsequent refinement network robustly removes the remaining artifacts.
The full network outperforms state-of-the-art reconstruction algorithms.
One of our core contributions is the discriminator perceptual network that enforces the network to incorporate high-level domain-specific knowledge.
We further introduce task-specific sub-modules tailored to the sparse-view tomography problems such as two-ends processing and the high-frequency loss.
Moreover, to test each module for effectiveness, we perform a comprehensive ablation study.

The two-step procedure showcases how to bring explainability to image reconstruction problems by tailoring the solution to the idiosyncrasies of the imaging environment.
We strongly believe that our approach is not limited to \gls{ct}, and we hope to inspire other imaging applications with similar ideas for pushing the limits of imaging.

{\small
\bibliographystyle{ieee_fullname}
\bibliography{paper}
}

\end{document}


\title{2-Step Sparse-View CT Reconstruction with a Domain-Specific Perceptual Network (Supplementary)}

\author{Haoyu Wei$^{1}$\thanks{The first two authors have equal contribution.}
\qquad
Florian Schiffers$^{1}$\footnotemark[1]
\qquad
Tobias Würfl$^{2}$
\qquad
Daming Shen$^{3}$\\
\qquad
Daniel Kim$^{3}$
\qquad
Aggelos Katsaggelos$^{1}$
\qquad
Oliver Cossairt$^{1}$\\
$^{1}$Northwestern University, Evanston, USA
\qquad
$^{2}$University of Erlangen-Nuremberg, Germany\\
\qquad
$^{3}$Feinberg School of Medicine, Northwestern University, Chicago, USA\\
\qquad
{\tt\small florian.schiffers@northwestern.edu}
}

\maketitle

\section{Introduction}
This document contains the supplementary materials that were left our in our main submission since they are too detailed for and do not significantly enhance the message we want to convey. Section~\ref{sec:dp} presents an incremental strategy for training GAN with a discriminator perceptual losses. Section~\ref{sec:obj} explains in detail the adversarial loss function mentioned in the main paper. The model architecture and hyperparameters are shown in Sec.~\ref{sec:detail}. Finally, more qualitative and quantitative results for ablation study are presented in Sec.~\ref{sec:ablation}.


We provide all code that is necessary to reproduce the experiments reported in the paper and supplementary at \url{https://github.com/anonyr7/Sinogram-Inpainting}.
This git-repository contains PyTorch implementations for our proposed networks, state-of-the-art methods, our CT dataset and a sub-set of around 30 images as examples for evaluation of each model.

\section{Discriminator Perceptual Loss}
\label{sec:dp}

We present an alternative training strategy using the DP loss in Alg.~\ref{algo:1}.
We show the example procedure with a super-resolution task that learns from a low-resolution (LR) image to a high-resolution (HR) image.
We start by training the discriminator once in every generator iteration, then increment the discriminator iteration by one in every $k$ generator iterations, where k is an arbitrary constant, for which we chose $k=10$ in our paper.
The intuition is that with more training performed, the discriminator learns better high-level features, thus benefits the generator using the DP loss.

\begin{algorithm}
\SetAlgoLined
\caption{An Incremental Training Procedure of a naive Super-Resolution GAN with DP loss.}
\KwData{LR and HR images; Generator training iteration $n$; Discriminator increment period $k$.}
\KwResult{Super-resolved (SR) images.}
 \While{not converge or iteration $i \leq n$}{
  Sample a batch of LR and HR image pairs $x$ and $y$ from the training data\;
  
  Generate a SR image $G(x)$ from generator $G$\;
  \While{discriminator iteration $j \leq \ceil[\big]{i/k}$ }{
    Update discriminator $D$ with $\mathcal{L}_{adv}(D)$\;
    j = j+1\;
  }
  Calculate $\mathcal{L}_{DP}$ from $D$ using $G(x)$ and $y$\;
  Update G with $\mathcal{L}_{DP}$ and $\mathcal{L}_{adv}(G)$\;
  i = i+1\;
 }
\label{algo:1}
\end{algorithm}

\section{Objective Function Details}
\label{sec:obj}
\subsection{Adversarial Loss}
For a typical GAN network that generates images from corrupted input $x$ to target distribution $y$, the adversarial losses for the generator and discriminator are defined as:
\begin{equation}
\begin{aligned}
\mathcal{L}_{adv}(G) =& \mathbb{E}_{x \sim corrupt} [\log (1-D(G(x)))], \\
\mathcal{L}_{adv}(D) =& \frac{1}{2}(\mathbb{E}_{y \sim target} [\log D(y)] + \\
& \mathbb{E}_{x \sim corrupt}[\log (1-D(G(x))]),
\end{aligned}
\label{equation:GAN}
\end{equation}
where $G$ and $D$ represent the generator and discriminator networks, respectively.

Equation \ref{equation:GAN} applies to the SIN with global discriminator and PRN model. For the local discriminator of SIN, a pair of image patches $G(x)_{patch}$ and $y_{patch}$ randomly selected from the generated image $G(x)$ and target image $y$ pair are used for calculating an additional local adversarial cost for SIN:
\begin{equation}
\begin{aligned}
\mathcal{L}_{local\_adv}(G) =& \mathbb{E}_{x \sim corrupt} [\log (1-D(G(x)_{patch}))], \\
\mathcal{L}_{local\_adv}(D) =& \frac{1}{2}(\mathbb{E}_{y \sim target} [\log D(y_{patch})] + \\
& \mathbb{E}_{x \sim corrupt}[\log (1-D(G(x)_{patch})]),
\end{aligned}
\label{equation:Local_GAN}
\end{equation}
Therefore, the total adversarial loss for SIN is the sum of Eq.~\ref{equation:GAN} and Eq.~\ref{equation:Local_GAN} with equal weights.




\section{More Training Details}
\label{sec:detail}

\subsection{Detailed Architecture}

As discussed in the main paper, our generator uses an adapted U-Net architecture, which consists of four average pooling and four bilinear upsampling layers, each scales by a factor of two in both dimensions. Each of the scaling layers is followed by a double-conv block, which consists of two sets of convolution, ReLU activation and batch normalization layers. All convolution layers in the U-Net have kernel size of three and stride equal to one.

We also use a patch discriminator~\cite{isola2017image}, where the first and last convolution layers have kernel size of three and stride of one, and the middle two convolutions have kernel size of four and stride of two. It also includes LeakyReLU activation and batch normalization layers. In total, the patch discriminator outputs a patch of size $1/4 \times 1/4$ than before.

Please find the implemention of this architecture here:
\url{https://github.com/anonyr7/Sinogram-Inpainting/blob/master/SIN/model.py}

\subsection{Data Augmentation}
We augment our dataset by applying random affine transformation to each image. Specifically, we randomly perform the following operations: rotate between $\pm30^{\circ}$, translate in two dimensions between $\pm0.1$, scale by $0.5$ to $1.1$, and shear between $\pm20^{\circ}$. Such augmented images are only used for training phase, in order to bring more diversity to both sinogram and reconstruction training sets.

The code for our data augmentation is found in this Jupyter notebook: 
\url{https://github.com/anonyr7/Sinogram-Inpainting/blob/master/TCIA_data_preprocessing.ipynb}

\section{More Results from Ablation Study and State-of-the-Art}
\label{sec:ablation}
In this section, we show complement of the ablation study discussed in the main paper. Specifically, we present more visual results of different composition of our models, analysis of the cascaded inputs and more sinogram-domain results for our discriminator perceptual loss.

\subsection{Checkerboard Artifacts in Sinograms}

\begin{figure*}
    \centering
    \begin{subfigure}[t]{.24\linewidth}
        \includegraphics[height=\linewidth]{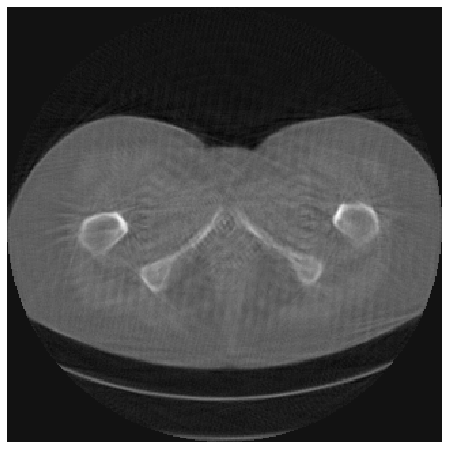}
    \end{subfigure}
    \begin{subfigure}[t]{.24\linewidth}
        \includegraphics[height=\linewidth]{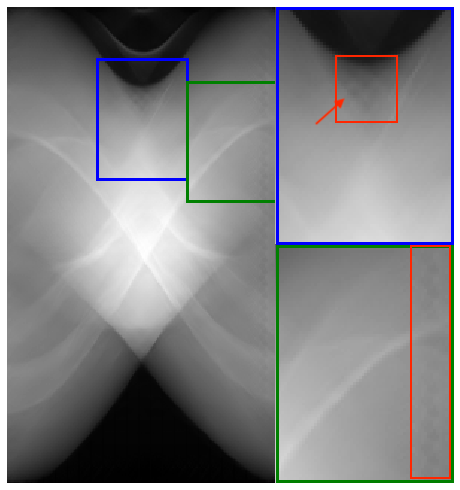}
    \end{subfigure}
    \begin{subfigure}[t]{.24\linewidth}
        \includegraphics[height=\linewidth]{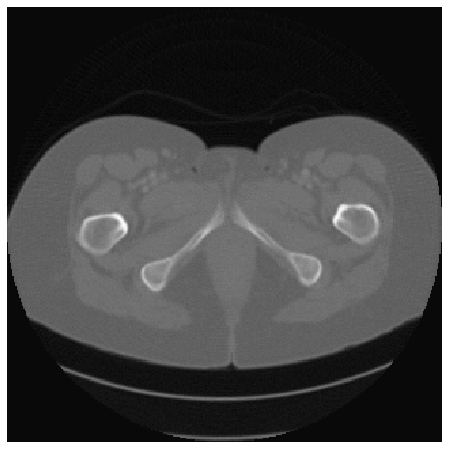}
    \end{subfigure}
    \begin{subfigure}[t]{.24\linewidth}
        \includegraphics[height=\linewidth]{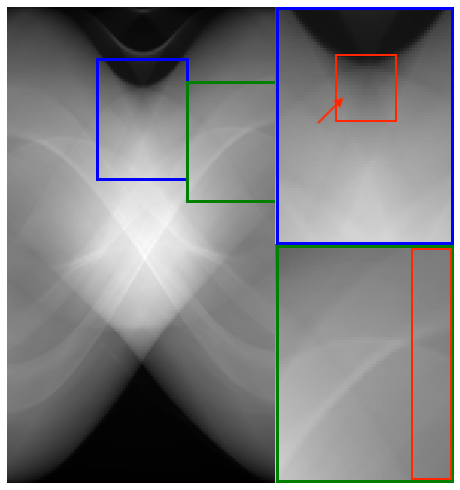}
    \end{subfigure}
    \begin{subfigure}[t]{.24\linewidth}
        \includegraphics[height=\linewidth]{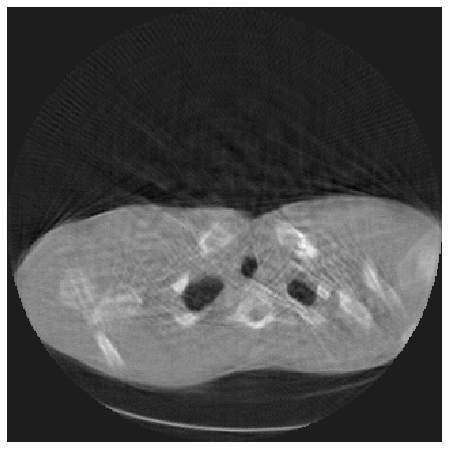}
    \end{subfigure}
    \begin{subfigure}[t]{.24\linewidth}
        \includegraphics[height=\linewidth]{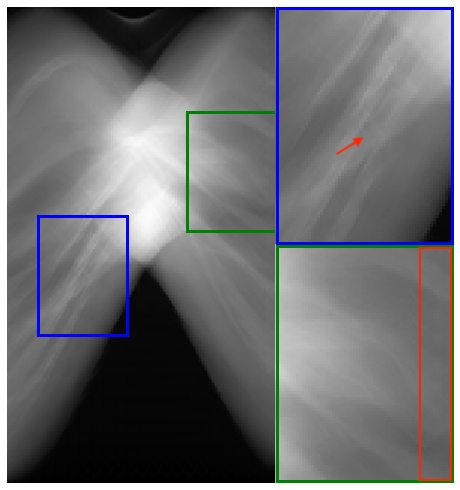}
    \end{subfigure}
    \begin{subfigure}[t]{.24\linewidth}
        \includegraphics[height=\linewidth]{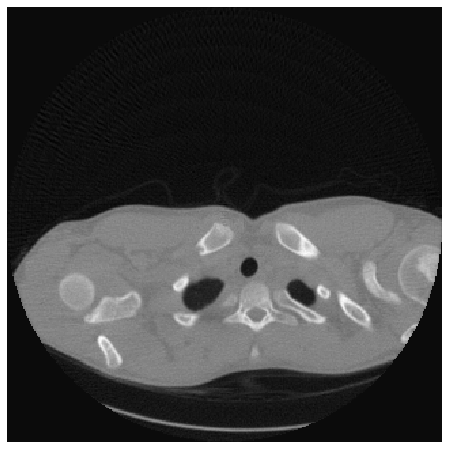}
    \end{subfigure}
    \begin{subfigure}[t]{.24\linewidth}
        \includegraphics[height=\linewidth]{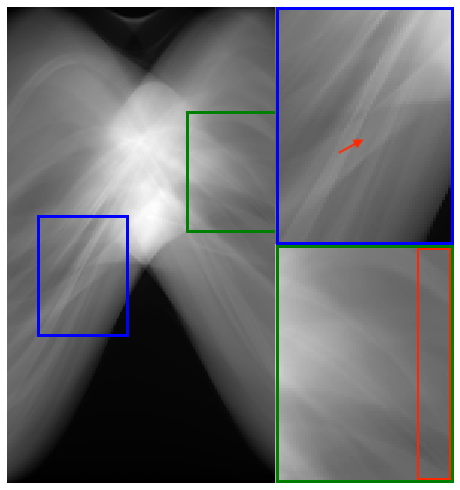}
    \end{subfigure}
    \caption{Checkerboard artifacts in the generated sinograms of the state-of-the-art learning-based approach~\cite{ghani2018deep} and consequent artifacts in their FBP reconstructions. \textbf{Left:} result sinograms and reconstructions. Checkerboard artifacts are especially severe in the boundary of sinograms. Please zoom in the sinogram to see more closely; \textbf{Right:} Ground Truth sinograms and reconstructions.}
    \label{fig:cgan}
\end{figure*}

We present more sinogram results from the cGAN model introduced by Ghani~\emph{et~al}~\cite{ghani2018deep} in Fig.~\ref{fig:cgan}. The sinograms exhibit streak artifacts especially on the boundary areas, causing the reconstructions to have skewed streak artifacts. More examples can be found at \url{https://github.com/anonyr7/Sinogram-Inpainting/tree/master/Toy-Dataset/CGAN_sinogram}.

\subsection{Qualitative comparison among our models}

\begin{figure*}[t]
\centering
    \begin{subfigure}[t]{.162\linewidth}
        \centering
        \begin{overpic}[width=\linewidth]{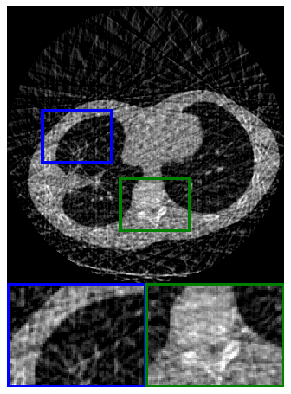}
         \put (5,90) {\textcolor{orange}{21.79}}
         \put (50,90) {\textcolor{orange}{0.446}} 
        \end{overpic}
    \end{subfigure}
    \begin{subfigure}[t]{.162\linewidth}
        \centering
        \begin{overpic}[width=\linewidth]{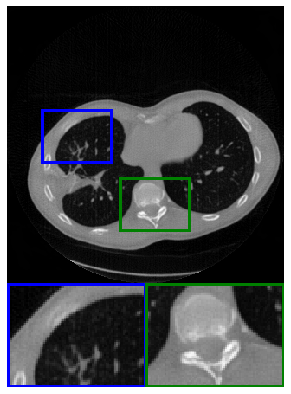}
         \put (5,90) {\textcolor{orange}{31.40}}
         \put (50,90) {\textcolor{orange}{0.810}}
        \end{overpic}
    \end{subfigure}
    \begin{subfigure}[t]{.162\linewidth}
        \centering
        \begin{overpic}[width=\linewidth]{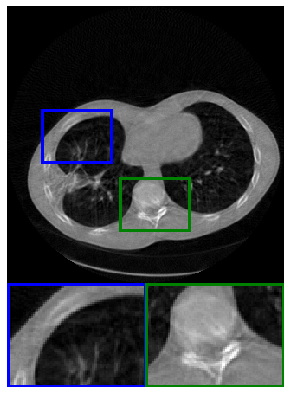}
         \put (5,90) {\textcolor{orange}{32.08}}
         \put (50,90) {\textcolor{orange}{0.831}}
        \end{overpic}
    \end{subfigure}
    \begin{subfigure}[t]{.162\linewidth}
        \centering
        \begin{overpic}[width=\linewidth]{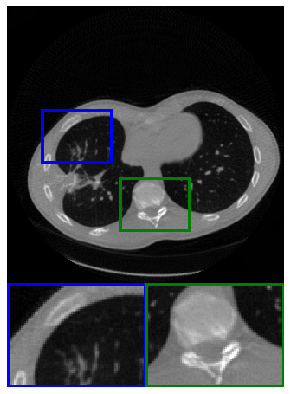}
         \put (5,90) {\textcolor{orange}{32.99}}
         \put (50,90) {\textcolor{orange}{0.858}}
        \end{overpic}
    \end{subfigure}
    \begin{subfigure}[t]{.162\linewidth}
        \centering
        \begin{overpic}[width=\linewidth]{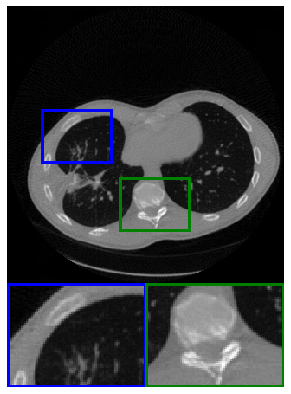}
         \put (5,90) {\textcolor{orange}{33.44}}
         \put (50,90) {\textcolor{orange}{0.865}}
        \end{overpic}
    \end{subfigure}
    \begin{subfigure}[t]{.162\linewidth}
        \centering
        \begin{overpic}[width=\linewidth]{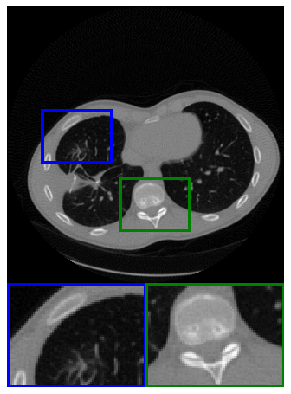}
         \put (5,90) {\textcolor{orange}{PSNR}}
         \put (50,90) {\textcolor{orange}{SSIM}}
        \end{overpic}
    \end{subfigure}
    \begin{subfigure}[t]{.162\linewidth}
        \centering
        \begin{overpic}[width=\linewidth]{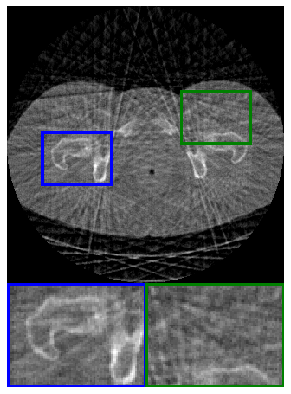}
         \put (5,90) {\textcolor{orange}{23.66}}
         \put (50,90) {\textcolor{orange}{0.515}}
        \end{overpic}
    \end{subfigure}
    \begin{subfigure}[t]{.162\linewidth}
        \centering
        \begin{overpic}[width=\linewidth]{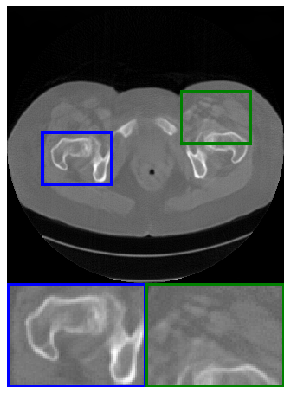}
         \put (5,90) {\textcolor{orange}{34.30}}
         \put (50,90) {\textcolor{orange}{0.869}}
        \end{overpic}
    \end{subfigure}
    \begin{subfigure}[t]{.162\linewidth}
        \centering
        \begin{overpic}[width=\linewidth]{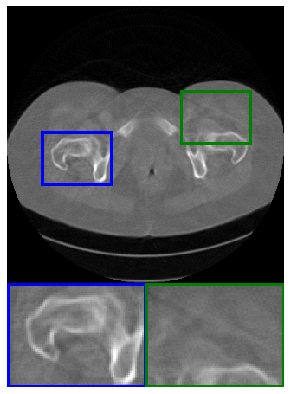}
         \put (5,90) {\textcolor{orange}{35.82}}
         \put (50,90) {\textcolor{orange}{0.887}}
        \end{overpic}
    \end{subfigure}
    \begin{subfigure}[t]{.162\linewidth}
        \centering
        \begin{overpic}[width=\linewidth]{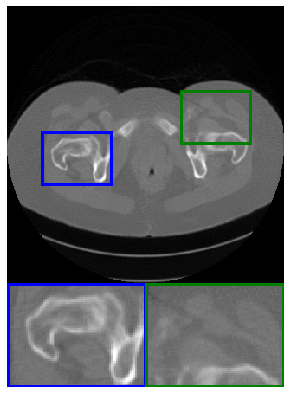}
         \put (5,90) {\textcolor{orange}{36.24}}
         \put (50,90) {\textcolor{orange}{0.898}}
        \end{overpic}
    \end{subfigure}
    \begin{subfigure}[t]{.162\linewidth}
        \centering
        \begin{overpic}[width=\linewidth]{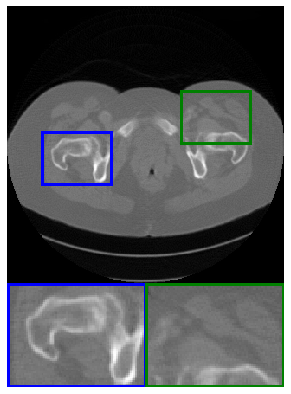}
         \put (5,90) {\textcolor{orange}{36.32}}
         \put (50,90) {\textcolor{orange}{0.903}}
        \end{overpic}
    \end{subfigure}
    \begin{subfigure}[t]{.162\linewidth}
        \centering
        \begin{overpic}[width=\linewidth]{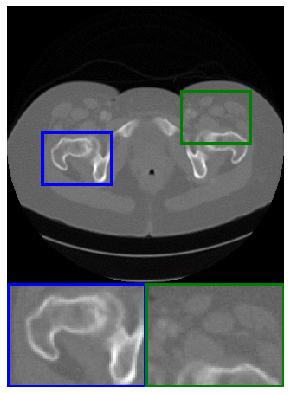}
         \put (5,90) {\textcolor{orange}{PSNR}}
         \put (50,90) {\textcolor{orange}{SSIM}}
        \end{overpic}
    \end{subfigure}
    \begin{subfigure}[t]{.162\linewidth}
        \centering
        \begin{overpic}[width=\linewidth]{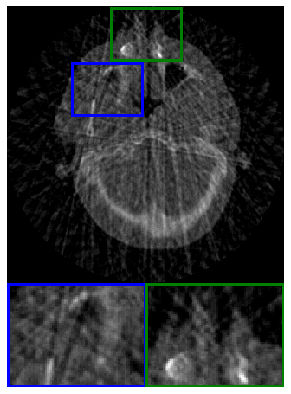}
         \put (5,90) {\textcolor{orange}{26.19}} 
         \put (50,90) {\textcolor{orange}{0.558}} 
        \end{overpic}
        \caption{Sparse-View}
    \end{subfigure}
    \begin{subfigure}[t]{.162\linewidth}
        \centering
        \begin{overpic}[width=\linewidth]{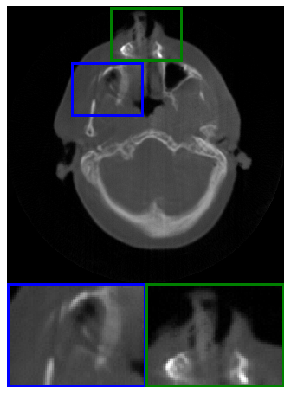}
         \put (5,90) {\textcolor{orange}{32.55}}
         \put (50,90) {\textcolor{orange}{0.887}}
        \end{overpic}
        \caption{PRN}
    \end{subfigure}
    \begin{subfigure}[t]{.162\linewidth}
        \centering
        \begin{overpic}[width=\linewidth]{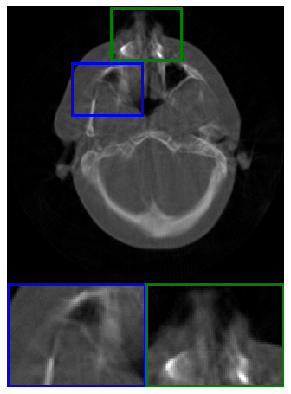}
         \put (5,90) {\textcolor{orange}{34.14}}
         \put (50,90) {\textcolor{orange}{0.894}}
        \end{overpic}
        \caption{SIN}
    \end{subfigure}
    \begin{subfigure}[t]{.162\linewidth}
        \centering
        \begin{overpic}[width=\linewidth]{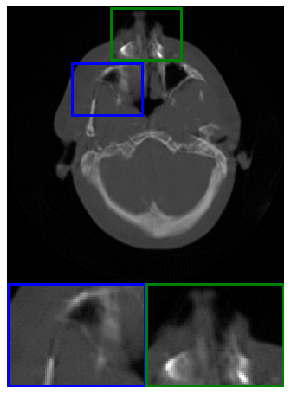}
         \put (5,90) {\textcolor{orange}{34.67}}
         \put (50,90) {\textcolor{orange}{0.920}}
        \end{overpic}
        \caption{SIN-PRN}
    \end{subfigure}
    \begin{subfigure}[t]{.162\linewidth}
        \centering
        \begin{overpic}[width=\linewidth]{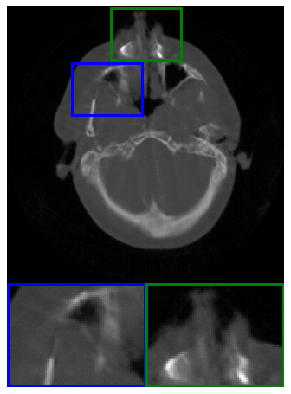}
         \put (5,90) {\textcolor{orange}{35.75}}
         \put (50,90) {\textcolor{orange}{0.919}}
        \end{overpic}
        \caption{SIN-4c-PRN}
    \end{subfigure}
    \begin{subfigure}[t]{.162\linewidth}
        \centering
        \begin{overpic}[width=\linewidth]{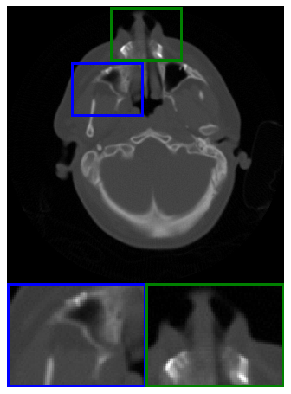}
         \put (5,90) {\textcolor{orange}{PSNR}}
         \put (50,90) {\textcolor{orange}{SSIM}}
        \end{overpic}
        \caption{Ground Truth}
    \end{subfigure}
    
    \caption{Ablation comparison with zoom-in details. From top to bottom row: 
    \textbf{Chest:} While PRN generates a visually clean image, it losses important details such as the bones in the blue box. In contrast, SIN-4c-PRN provides both smooth and detail-preserved reconstructions.
    \textbf{Abdomen:} PRN fails to remove the heavy streak artifacts in the green box area. 
    \textbf{Head:} PRN hallucinates non-exist features on the nose and eye regions, while SIN-4c-PRN produces results close to the ground truth.
    }
    \label{fig:models}
\end{figure*}

We present the visual results of different compositions of our models in Fig.~\ref{fig:models}, corresponding to the quantitative metrics among our models in the main paper.
We show that although a single PRN model generates smooth and clean images, they still either suffer from unremoved streak artifacts or hallucinate wrong details that are critical in analysis of the reconstructions.
Meanwhile, we also observe secondary artifacts from single SIN model.
This is in correspondence with our argument that single-domain models are not enough for SV-CT reconstruction tasks, and our SIN-4c-PRN model benefits from both domain representations.

\subsection{Effectiveness of cascade inputs of PRN}
\label{sec:4c}
Our proposed model SIN-4c-PRN is a concatenation of SIN and PRN models trained in two steps. In particular, we propose a four-channel input for PRN that consists of reconstructions from sinograms with different undersampling factors. 

Specifically, we create an input with each channel being a FBP reconstruction from a sinogram with 23, 45, 90 and 180 angles, respectively. The 23-angle sinogram is the original sparse-view sinogram, and the intermediate angles are downsampled from the 180-angle SIN-inpainted sinogram.

In order to make fair comparisons of the proposed cascade inputs, we train two networks with the same capacity, i.e., number of learnable parameters. SIN-4x-PRN is a model with a repeated four-channel input, each channel being the same FBP reconstruction from the 180-angle SIN-inpainted sinogram. We compare the performance of SIN-4x-PRN with SIN-4c-PRN in Tb.~\ref{table:metric}. Both PSNR and SSIM metrics show that SIN-4c-PRN performs better than SIN-4x-PRN, indicating that the model learns from the cascaded inputs with different level of representations.

\begin{table}
\centering
\begin{tabularx}{\linewidth}{|X|c|c|}
\hline
Model & PSNR ($\sigma$) & SSIM ($\sigma$) \\
\hline\hline
SIN-PRN & 34.61 (2.14) & 0.873 (0.035)\\
SIN-4x-PRN & 34.62 (1.97) & 0.874 (0.029) \\
SIN-4c-PRN & \textbf{34.90 (2.15)} & \textbf{0.877 (0.029)}\\
\hline
\end{tabularx}
\caption{Reconstruction performance comparison. SIN-4x-PRN: Same model with SIN-4c-RPN, but duplicating input reconstruction by four times. $\sigma$ denotes standard deviation.}
\label{table:metric}
\end{table}

\begin{table}
\centering
\begin{tabularx}{\linewidth}{|c|X|c|c|}
    \hline
     & Model & PSNR ($\sigma$) & SSIM ($\sigma$) \\
    \hline\hline
    \multirow{3}{*}{S} & Non-Perceptual & 42.50 (2.54) & 0.979 (0.005)\\
    & VGG16~\cite{johnson2016perceptual} & 39.70 (2.25) & 0.984 (0.004) \\
    & Ours (DP) & 41.23 (2.83) & 0.988 (0.005)\\
    \hline
    \multirow{3}{*}{R} & Non-Perceptual &33.39 (2.09) &0.829 (0.037)\\
    & VGG16~\cite{johnson2016perceptual} &33.51 (2.02) &0.852 (0.036) \\
    & Ours (DP) & 34.19 (2.35) & 0.859 (0.036)\\
    \hline
\end{tabularx}
\caption{Different perceptual loss performance trained with a single SIN. `S' and `R' denote Sinogram and Reconstruction domain, respectively. $\sigma$ denotes standard deviation.}
\label{table:sino_metric}
\end{table}

\subsection{Sinogram results for different perceptual losses}
\label{sec:dp_result}

We qualitatively and quantitatively compare the sinogram results of our DP loss with the original VGG16 perceptual loss~\cite{johnson2016perceptual} trained with a SIN model. The SIN model learns 1D super-resolution from a 23-angle sparse-view sinogram to a 180-angle full-view sinogram. The models used for comparison share the same architecture, data and training procedure, thus all the differences are contributed by the different perceptual losses.

In the main paper we compared the PSNR and SSIM metrics of the FBP reconstructions of result sinograms.
Here, we further provide metrics measured directly on the result sinograms in Tb.~\ref{table:sino_metric}. We observe better metrics of DP than VGG16, despite both of them have slightly lower PSNR than without perceptual loss. This is consistent with the observation in the original work by Johnson~\emph{et~al.}~\cite{johnson2016perceptual}, because perceptual losses are optimized with a different criterion, i.e., the $l_2$-norm is minimal when one optimizes for the $l_2$ loss.

Then we show the visual results of the sinograms and their FBP reconstructions in Fig.~\ref{fig:perceptual}. Specifically, we provide the sinogram residuals compared to the ground truth for ease of comparison. From these residuals, we find that VGG16 produce slightly more errors than the others in the sinogram domain, i.e., the training domain, possibly because sinograms are drastically different from natural RGB images where VGG16 was trained with. However, we also observe that after the FBP reconstruction, VGG16 and DP produce visually less noisy images than without perceptual losses. The reconstruction domain metrics and residuals indicate the same observation.

\begin{figure*}[h]
\centering
    \begin{subfigure}[t]{.24\linewidth}
        \centering
        \begin{overpic}[width=\linewidth]{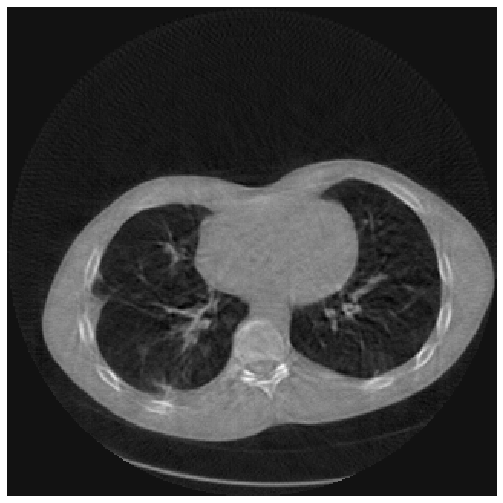}
         \put (7,87) {\textcolor{orange}{31.30}}
         \put (72,87) {\textcolor{orange}{0.806}} 
        \end{overpic}
    \end{subfigure}
    \begin{subfigure}[t]{.24\linewidth}
        \centering
        \begin{overpic}[width=\linewidth]{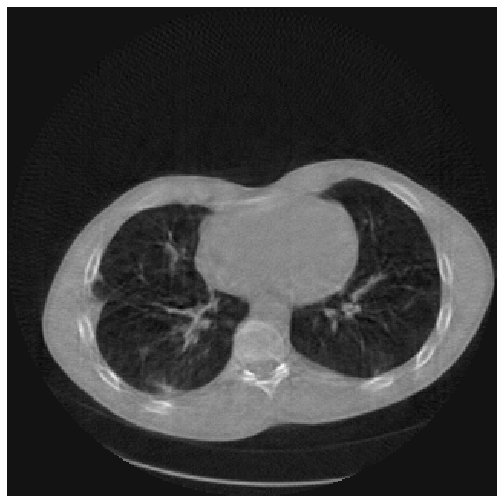}
         \put (7,87) {\textcolor{orange}{31.47}}
         \put (72,87) {\textcolor{orange}{0.823}}
        \end{overpic}
    \end{subfigure}
    \begin{subfigure}[t]{.24\linewidth}
        \centering
        \begin{overpic}[width=\linewidth]{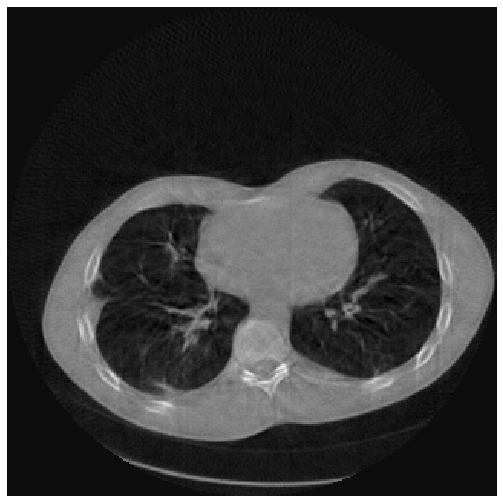}
         \put (7,87) {\textcolor{orange}{31.96}}
         \put (72,87) {\textcolor{orange}{0.828}}
        \end{overpic}
    \end{subfigure}
    \begin{subfigure}[t]{.24\linewidth}
        \centering
        \begin{overpic}[width=\linewidth]{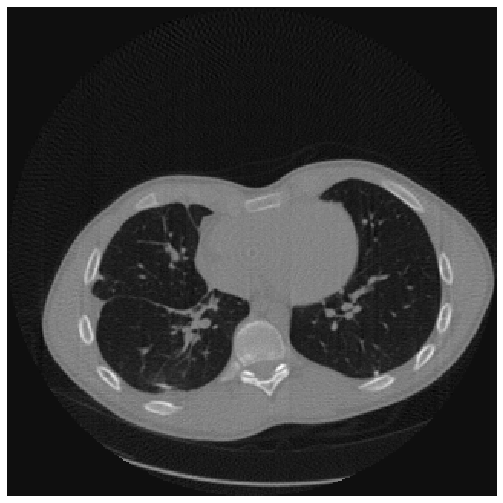}
         \put (7,87) {\textcolor{orange}{PSNR}}
         \put (72,87) {\textcolor{orange}{SSIM}}
        \end{overpic}
    \end{subfigure}
    \begin{subfigure}[t]{.24\linewidth}
        \centering
        \begin{overpic}[width=\linewidth]{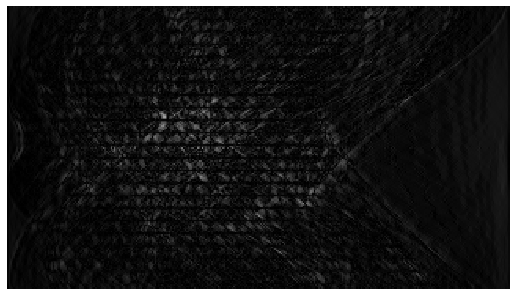}
        \end{overpic}
    \end{subfigure}
    \begin{subfigure}[t]{.24\linewidth}
        \centering
        \begin{overpic}[width=\linewidth]{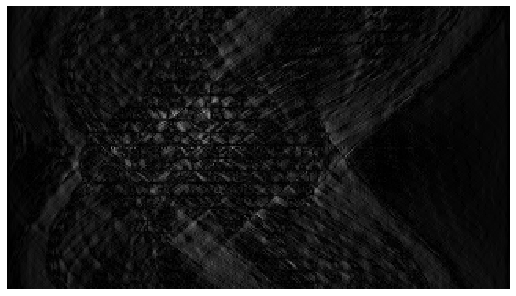}
        \end{overpic}
    \end{subfigure}
    \begin{subfigure}[t]{.24\linewidth}
        \centering
        \begin{overpic}[width=\linewidth]{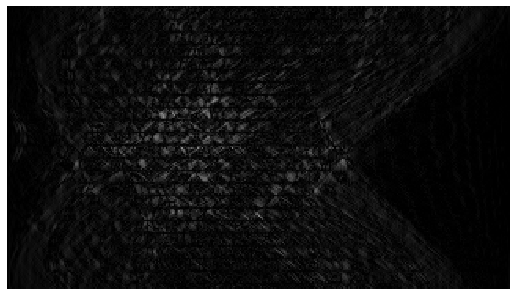}
        \end{overpic}
    \end{subfigure}
    \begin{subfigure}[t]{.24\linewidth}
        \centering
        \begin{overpic}[width=\linewidth]{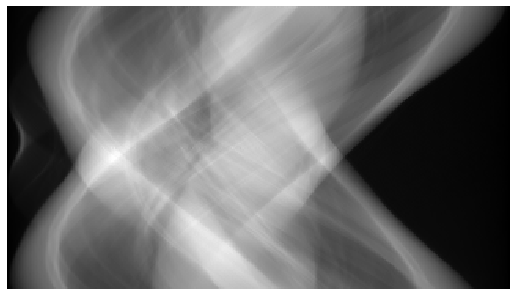}
        \end{overpic}
    \end{subfigure}
    \begin{subfigure}[t]{.24\linewidth}
        \centering
        \begin{overpic}[width=\linewidth]{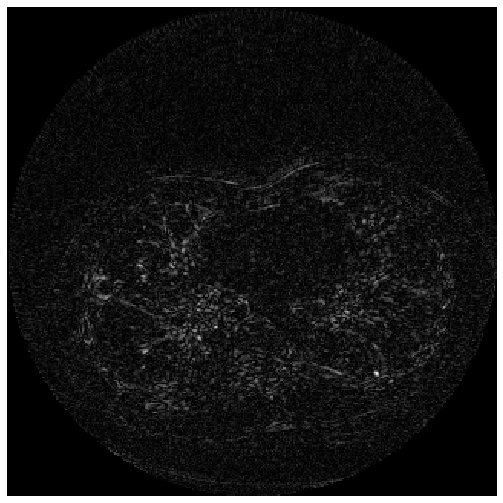}
        \end{overpic}
    \end{subfigure}
    \begin{subfigure}[t]{.24\linewidth}
        \centering
        \begin{overpic}[width=\linewidth]{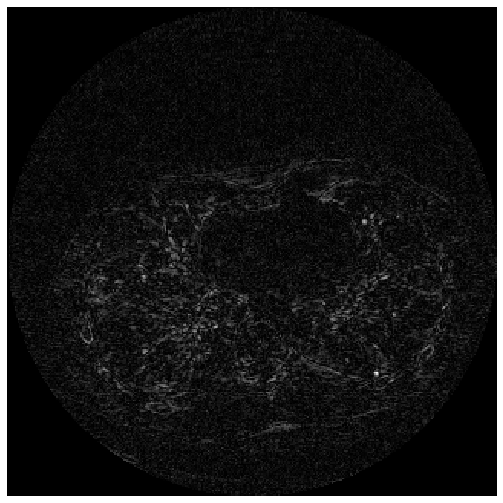}
        \end{overpic}
    \end{subfigure}
    \begin{subfigure}[t]{.24\linewidth}
        \centering
        \begin{overpic}[width=\linewidth]{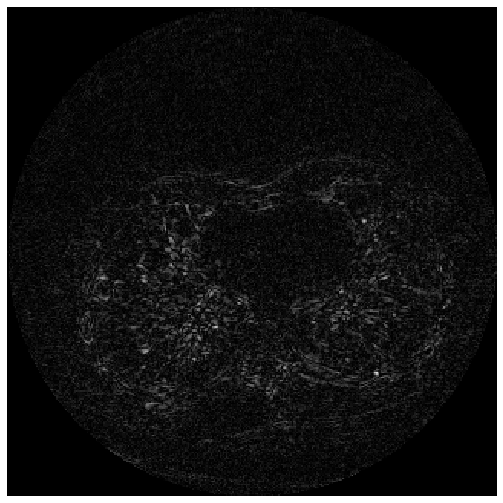}
        \end{overpic}
    \end{subfigure}
    \begin{subfigure}[t]{.24\linewidth}
        \centering
        \begin{overpic}[width=\linewidth]{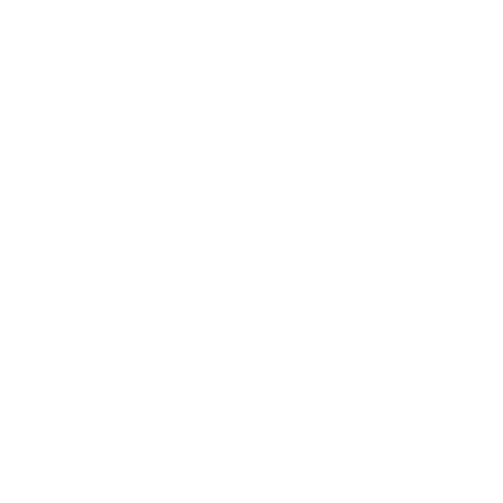}
        \put (40,50) {\color{black}\large{\textbf{Chest}}}
        \end{overpic}
    \end{subfigure}
    \begin{subfigure}[t]{.24\linewidth}
        \centering
        \begin{overpic}[width=\linewidth]{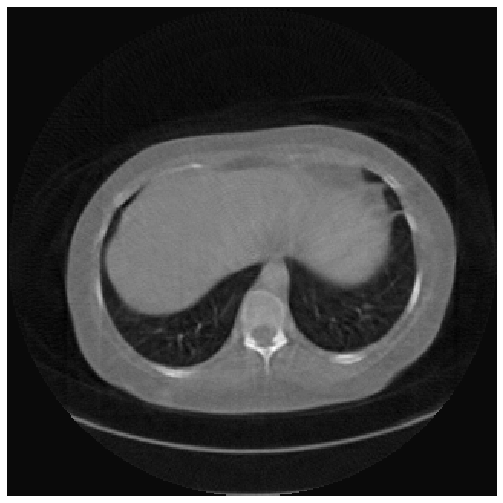}
         \put (7,87) {\textcolor{orange}{35.73}}
         \put (72,87) {\textcolor{orange}{0.885}}
        \end{overpic}
    \end{subfigure}
    \begin{subfigure}[t]{.24\linewidth}
        \centering
        \begin{overpic}[width=\linewidth]{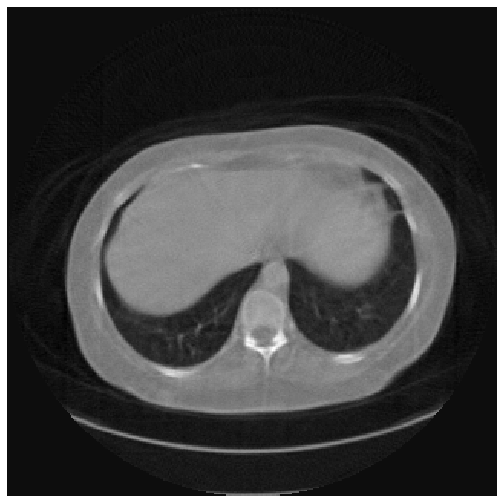}
         \put (7,87) {\textcolor{orange}{35.04}}
         \put (72,87) {\textcolor{orange}{0.897}}
        \end{overpic}
    \end{subfigure}
    \begin{subfigure}[t]{.24\linewidth}
        \centering
        \begin{overpic}[width=\linewidth]{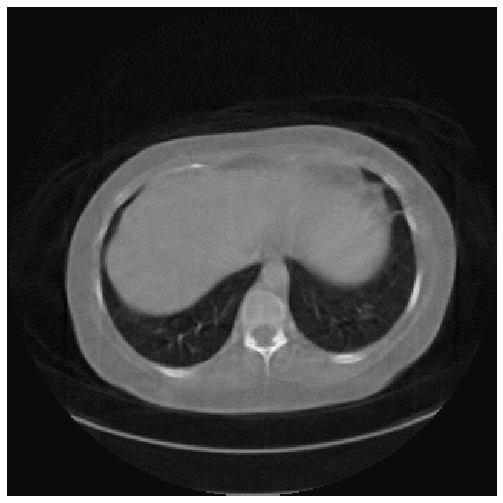}
         \put (7,87) {\textcolor{orange}{36.45}}
         \put (72,87) {\textcolor{orange}{0.908}}
        \end{overpic}
    \end{subfigure}
    \begin{subfigure}[t]{.24\linewidth}
        \centering
        \begin{overpic}[width=\linewidth]{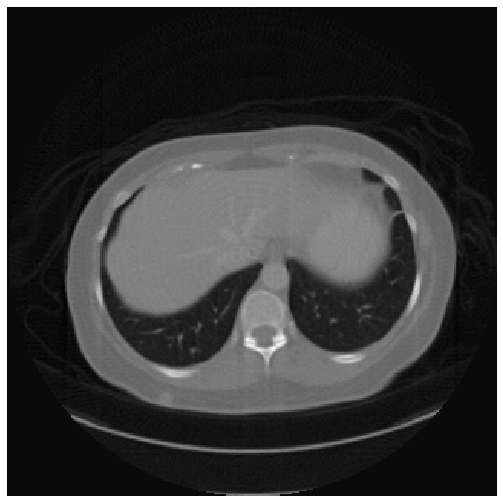}
         \put (7,87) {\textcolor{orange}{PSNR}}
         \put (72,87) {\textcolor{orange}{SSIM}}
        \end{overpic}
    \end{subfigure}
    \begin{subfigure}[t]{.24\linewidth}
        \centering
        \begin{overpic}[width=\linewidth]{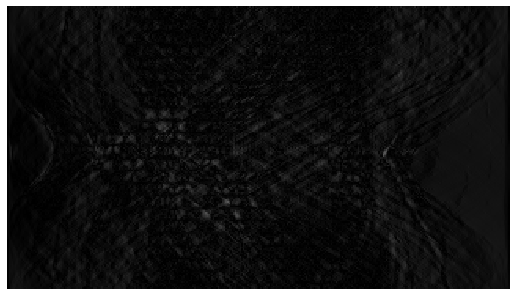}
        \end{overpic}
    \end{subfigure}
    \begin{subfigure}[t]{.24\linewidth}
        \centering
        \begin{overpic}[width=\linewidth]{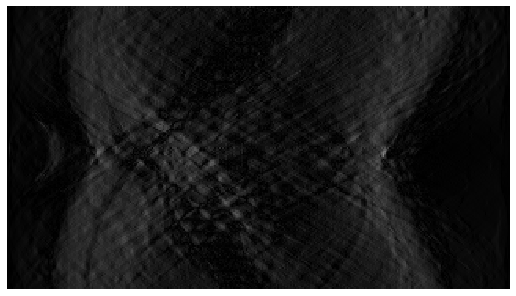}
        \end{overpic}
    \end{subfigure}
    \begin{subfigure}[t]{.24\linewidth}
        \centering
        \begin{overpic}[width=\linewidth]{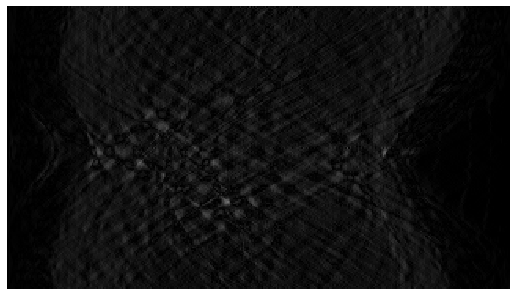}
        \end{overpic}
    \end{subfigure}
    \begin{subfigure}[t]{.24\linewidth}
        \centering
        \begin{overpic}[width=\linewidth]{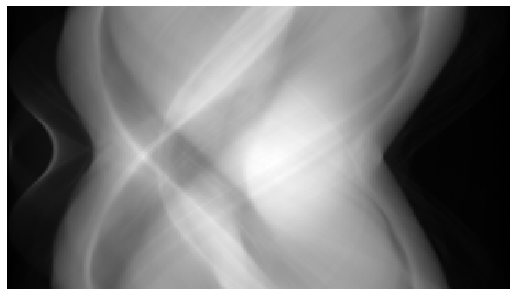}
        \end{overpic}
    \end{subfigure}
    \begin{subfigure}[t]{.24\linewidth}
        \centering
        \begin{overpic}[width=\linewidth]{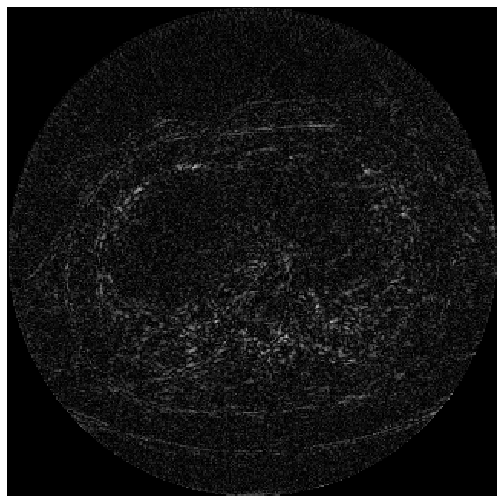}
        \end{overpic}
    \end{subfigure}
    \begin{subfigure}[t]{.24\linewidth}
        \centering
        \begin{overpic}[width=\linewidth]{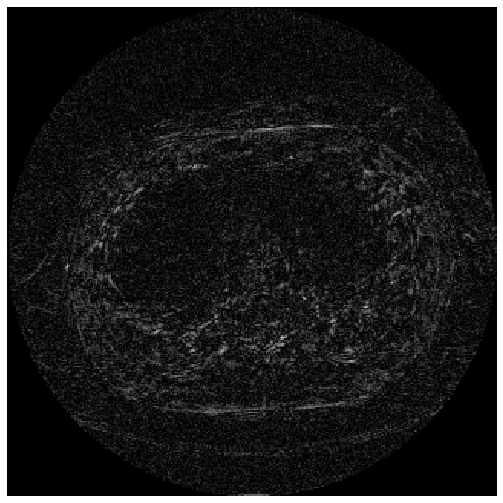}
        \end{overpic}
    \end{subfigure}
    \begin{subfigure}[t]{.24\linewidth}
        \centering
        \begin{overpic}[width=\linewidth]{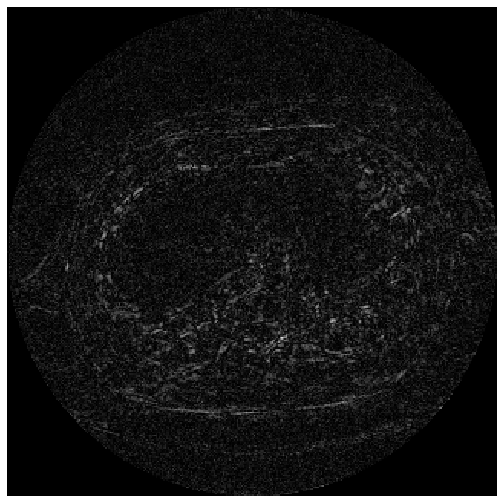}
        \end{overpic}
    \end{subfigure}
    \begin{subfigure}[t]{.24\linewidth}
        \centering
        \begin{overpic}[width=\linewidth]{images/empty.png}
        \put (30,50) {\color{black}\large{\textbf{Abdomen}}}
        \end{overpic}
    \end{subfigure}
\end{figure*}

\begin{figure*}[h]
\ContinuedFloat
\centering
    \begin{subfigure}[t]{.24\linewidth}
        \centering
        \begin{overpic}[width=\linewidth]{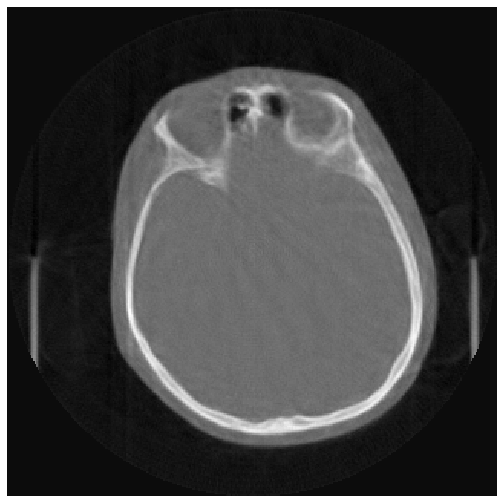}
         \put (7,87) {\textcolor{orange}{35.89}}
         \put (72,87) {\textcolor{orange}{0.889}}
        \end{overpic}
    \end{subfigure}
    \begin{subfigure}[t]{.24\linewidth}
        \centering
        \begin{overpic}[width=\linewidth]{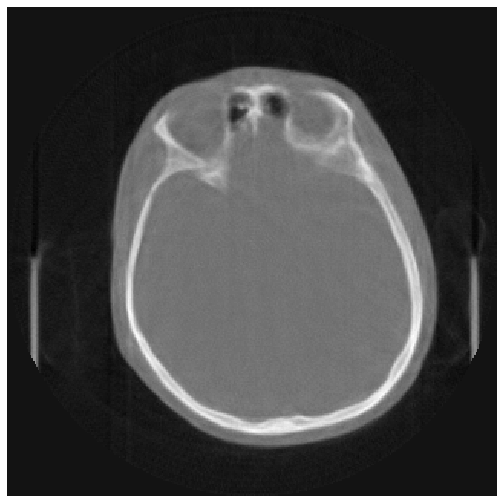}
         \put (7,87) {\textcolor{orange}{36.22}}
         \put (72,87) {\textcolor{orange}{0.917}}
        \end{overpic}
    \end{subfigure}
    \begin{subfigure}[t]{.24\linewidth}
        \centering
        \begin{overpic}[width=\linewidth]{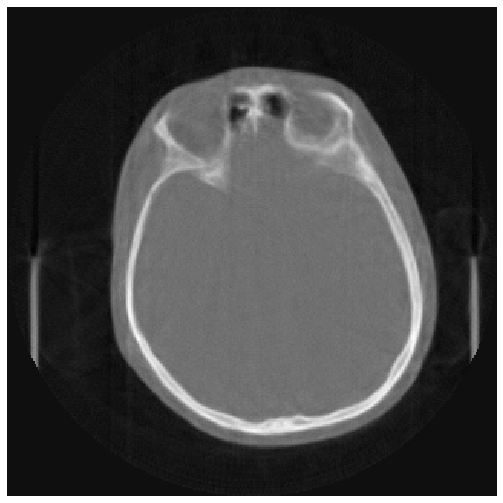}
         \put (7,87) {\textcolor{orange}{37.08}}
         \put (72,87) {\textcolor{orange}{0.922}}
        \end{overpic}
    \end{subfigure}
    \begin{subfigure}[t]{.24\linewidth}
        \centering
        \begin{overpic}[width=\linewidth]{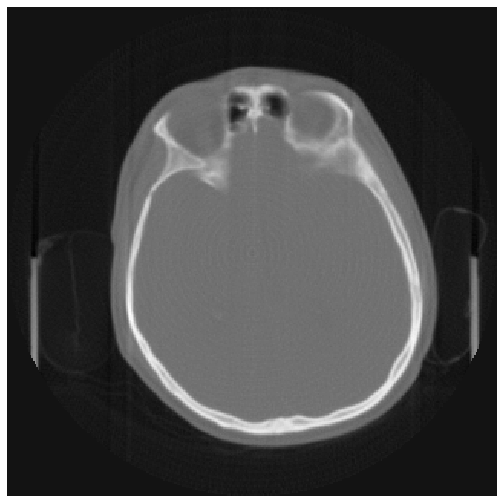}
         \put (7,87) {\textcolor{orange}{PSNR}}
         \put (72,87) {\textcolor{orange}{SSIM}}
        \end{overpic}
    \end{subfigure}
    \begin{subfigure}[t]{.24\linewidth}
        \centering
        \begin{overpic}[width=\linewidth]{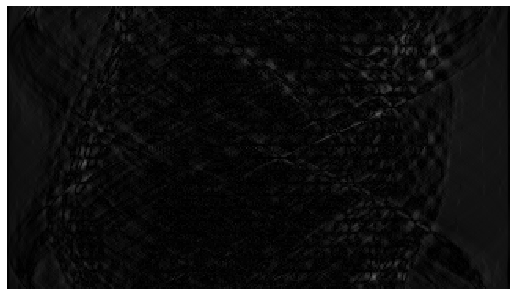}
        \end{overpic}
    \end{subfigure}
    \begin{subfigure}[t]{.24\linewidth}
        \centering
        \begin{overpic}[width=\linewidth]{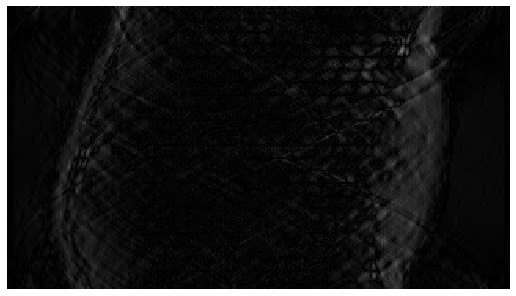}
        \end{overpic}
    \end{subfigure}
    \begin{subfigure}[t]{.24\linewidth}
        \centering
        \begin{overpic}[width=\linewidth]{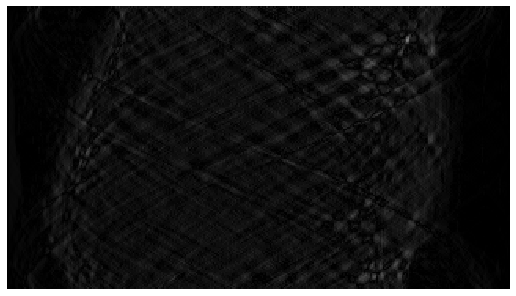}
        \end{overpic}
    \end{subfigure}
    \begin{subfigure}[t]{.24\linewidth}
        \centering
        \begin{overpic}[width=\linewidth]{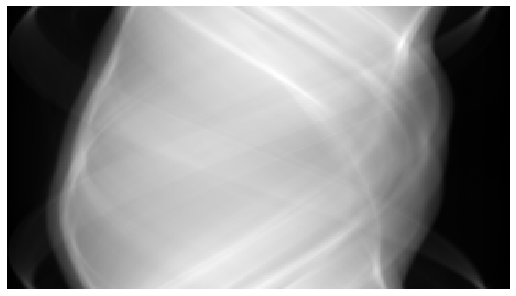}
        \end{overpic}
    \end{subfigure}
    \begin{subfigure}[t]{.24\linewidth}
        \centering
        \begin{overpic}[width=\linewidth]{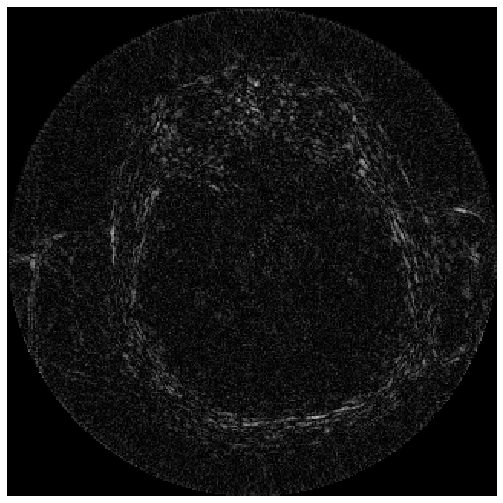}
        \end{overpic}
        \caption{Non-Perceptual}
    \end{subfigure}
    \begin{subfigure}[t]{.24\linewidth}
        \centering
        \begin{overpic}[width=\linewidth]{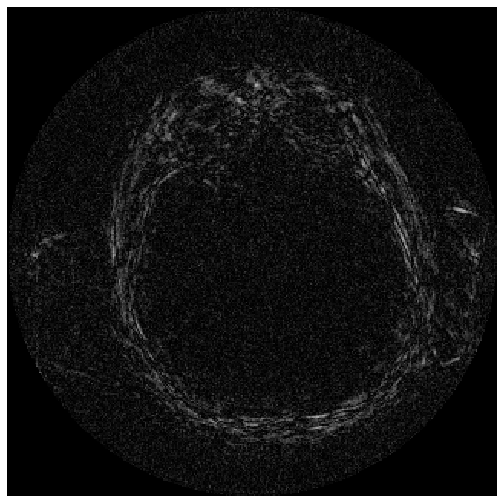}
        \end{overpic}
        \caption{VGG16~\cite{johnson2016perceptual}}
    \end{subfigure}
    \begin{subfigure}[t]{.24\linewidth}
        \centering
        \begin{overpic}[width=\linewidth]{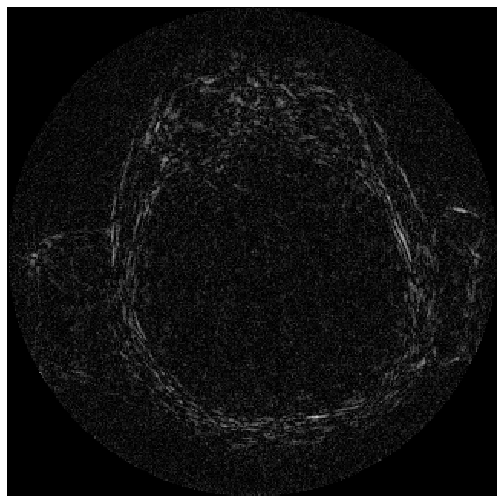}
        \end{overpic}
        \caption{Ours (DP)}
    \end{subfigure}
    \begin{subfigure}[t]{.24\linewidth}
        \centering
        \begin{overpic}[width=\linewidth]{images/empty.png}
        \put (40,50) {\color{black}\large{\textbf{Head}}}
        \end{overpic}
        \caption{Ground Truth}
    \end{subfigure}
    \caption{ (Continued) Sinogram residuals and corresponding FBP reconstructions and residuals from a single SIN using no perceptual loss, VGG16 perceptual loss or our discriminator perceptual loss. The residual (error) maps are displayed in the same range, hence we omit the scale-bar for visualization reasons. Both the non-perceptual and our DP generated sinograms have fewer errors than VGG16. However, the reconstructions of Non-perceptual sinograms are slightly noiser and have more remaining streak artifacts than the others. 
    }
    \label{fig:perceptual}
\end{figure*}

{\small
\bibliographystyle{ieee_fullname}
\bibliography{supp}
}